\documentclass[prl,aps,floats,twocolumn,superscriptaddress,floatfix,longbibliography]{revtex4-2}
\usepackage{amssymb,amsmath}

\usepackage[mathscr]{euscript}
\usepackage{graphicx,cases}
\usepackage{epsfig}
\usepackage{textcomp}
\usepackage{epstopdf}
\usepackage{float}
\usepackage[normalem]{ulem}
\usepackage{enumerate}
\usepackage{enumitem}
\usepackage{psfrag}
\usepackage[makeroom]{cancel}
\usepackage{xcolor}
\usepackage{footmisc}
\usepackage{soul}
\usepackage[%
  colorlinks=true,
  urlcolor=blue,
  linkcolor=blue,
  citecolor=blue
]{hyperref}

\def\beq{\begin{equation}}
\def\eeq{\end{equation}}
\def\bea{\begin{eqnarray}}
\def\eea{\end{eqnarray}}

\definecolor{mygreen}{rgb}{0.0,0.55,0.3}

\DeclareMathOperator{\sinc}{sinc}
\begin{document}

 \title{Self-organized hyperuniformity in a minimal model of population dynamics}
\author{  Tal Agranov}
\affiliation{Department of Applied Mathematics and Theoretical Physics, Centre for Mathematical Sciences, University of Cambridge, Cambridge CB3 0WA, UK}
\affiliation{Gurdon Institute, University of Cambridge, Cambridge CB2 1QN, UK}
\author{Natan Wiegenfeld}
\affiliation{Department of Applied Mathematics and Theoretical Physics, Centre for Mathematical Sciences, University of Cambridge, Cambridge CB3 0WA, UK}
\affiliation{Cavendish Laboratory, University of Cambridge, Cambridge CB3 0US, UK}
\author{Omer Karin}
\affiliation{Department of Mathematics, Imperial College London, London, SW7 2AZ, UK}
\author{Benjamin D. Simons}
\affiliation{Department of Applied Mathematics and Theoretical Physics, Centre for Mathematical Sciences, University of Cambridge, Cambridge CB3 0WA, UK}
\affiliation{Gurdon Institute, University of Cambridge, Cambridge CB2 1QN, UK}

\begin{abstract}
\noindent By generalizing a class of models recently introduced to account for protracted transients in biological systems, we identify a novel mechanism for hyperuniformity. In this model, competition of individuals over a shared resource serves as feedback that can asymptotically guide the population towards a critical steady state with divergent individual life time. We show that, in its spatially extended form, this many-body model exhibits hyperuniform density fluctuations. Through explicit coarse-graining, we develop a hydrodynamic theory that conforms closely with the results of stochastic simulations. Unlike previous models for non-equilibrium hyperuniform states, our model does not exhibit conservation laws, even in the asymptotic regime. Instead, hyperuniformity arises from the divergence of the range of the resource-mediated interactions. These findings may find applications in engineering, cellular population dynamics, and ecology.
 \end{abstract} 

\maketitle

\noindent \emph{Introduction} -- 
Over the past two decades, the phenomenon of hyperuniformity has drawn increasing attention across multiple contexts and
disciplines~\cite{torquatoLocalDensityFluctuations2003,torquatoHyperuniformStatesMatter2018,leiNonequilibriumDynamicHyperuniform2025}. It represents the remarkable property whereby a system of particles is statistically disordered at short length scales, yet has vanishing density fluctuations at large scales, as in zero-temperature crystals~\cite{torquatoHyperuniformStatesMatter2018}. Formally, the variance of the number of particles $N(\ell)$ in a given domain of size $\ell$ scales sub-extensively with volume, defying generic central limit theorem arguments. Mathematically, this translates to the condition
\begin{eqnarray}\label{hyperuniformity}
\frac{\text{Var}\left[N(\ell)\right]}{\langle N(\ell)\rangle}\sim \ell^{-\alpha},
\end{eqnarray} 
where the exponent $0<\alpha\leq1$ characterizes the degree of hyperuniformity~\cite{torquatoHyperuniformStatesMatter2018}. Equivalently, hyperuniformity can be reformulated as the vanishing of the structure factor, $\lim_{q\to0}S\left(q\right)=\lim_{\ell\to\infty}\text{Var}\left[N(\ell)\right]/\langle N(\ell)\rangle\stackrel{!}{=}0$ \cite{torquatoLocalDensityFluctuations2003}, with
\begin{eqnarray}\label{sofq}
S(q)\equiv\frac{\langle\delta\rho(\mathbf{q})\delta\rho(-\mathbf{q})\rangle}{\langle\rho\rangle},
\end{eqnarray}
where $\delta\rho(\mathbf{q})$ denote density fluctuations around the mean $\langle\rho\rangle$ at Fourier mode $\mathbf{q}$ and $S$ is a function of $q=|\mathbf{q}|$ from isotropy. Realizations of hyperuniformity have been established in several synthetic settings, including athermally jammed hard spheres~\cite{donevUnexpectedDensityFluctuations2005,zacharyHyperuniformLongRangeCorrelations2011}, periodically driven emulsions~\cite{tjhungHyperuniformDensityFluctuations2015,weijsEmergentHyperuniformityPeriodically2015}, and active Quincke rollers~\cite{zhangHyperuniformActiveChiral2022}. More recently, biologically occurring instances of hyperuniformity have been identified, including the distribution of avian photoreceptors~\cite{kramAvianConePhotoreceptors2010,jiaoAvianPhotoreceptorPatterns2014}, the leaf vein network~\cite{liuUniversalHyperuniformOrganization2024}, and vegetation coverage in semi-arid landscapes~\cite{geHiddenOrderTuring2023,huDisorderedHyperuniformitySignals2023,ezoeWeightedPointConfigurations2025}. In a solid state setting, interest in hyperuniformity draws from their remarkable material properties, such as complete photonic band gaps~\cite{florescuDesignerDisorderedMaterials2009,manIsotropicBandGaps2013,froufe-perezRoleShortRangeOrder2016,leseurHighdensityHyperuniformMaterials2016,liBiologicalTissueinspiredTunable2018} and superior mechanical ~\cite{zhangPerfectGlassParadigm2016,xuMicrostructureMechanicalProperties2017,chenDesigningDisorderedHyperuniform2018} and acoustic~\cite{gkantzounisHyperuniformDisorderedPhononic2017,cheronExperimentalCharacterizationRigidscatterer2022} properties. In biological contexts, hyperuniformity may serve important functions, such as providing uniform coverage of cell types~\cite{kramAvianConePhotoreceptors2010,jiaoAvianPhotoreceptorPatterns2014} or the optimization of resource-acquisition~\cite{geHiddenOrderTuring2023,huDisorderedHyperuniformitySignals2023} and resource-distribution~\cite{liuUniversalHyperuniformOrganization2024}.
 
Finding generic mechanisms that give rise to hyperuniformity remains an outstanding challenge which, if met, can unlock the potential for novel functional materials and advance our understanding of the mechanisms driving hyperuniformity in biological systems. While equilibrium dynamics with short-ranged interactions cannot result in hyperuniformity~\cite{kimEffectImperfectionsHyperuniformity2018}, non-equilibrium systems, such as the examples listed above, can relax to a hyperuniform state~\cite{leiNonequilibriumDynamicHyperuniform2025}. Prominent models that capture such non-equilibrium hyperuniformity include those with an absorbing state phase transition at criticality~\cite{maTheoryHyperuniformityAbsorbing2023,corteSelfOrganizedCriticalitySheared2009,hexnerNoiseDiffusionHyperuniformity2017,hexnerHyperuniformityCriticalAbsorbing2015,wieseHyperuniformityMannaModel2024}, active and passive phase separating systems during spinodal decomposition~\cite{delucaHyperuniformityPhaseOrdering2024}, and various active fluid models~\cite{leiHydrodynamicsRandomorganizingHyperuniform2019,leiNonequilibriumStronglyHyperuniform2019,zhangHyperuniformActiveChiral2022,cengioGiantDensityFluctuations2024}. In these contexts, steady state hyperuniformity emerges when a control parameter is fine-tuned to a critical value for a phase transition. A unifying feature that underlies hyperuniformity in many of these systems is the effective emergence of center-of-mass conservation at criticality~\cite{hexnerNoiseDiffusionHyperuniformity2017,bertrandNonlinearDiffusionHyperuniformity2019,delucaHyperuniformityPhaseOrdering2024,maHyperuniformityAbsorbingState2025,maTheoryHyperuniformityAbsorbing2023,cengioGiantDensityFluctuations2024,mukherjeeAnomalousRelaxationHyperuniform2024,hazraHyperuniformityMassTransport2025,maireHyperuniformityConservationLaws2025}. This is manifest in ‘superconservative’ noise whose amplitude vanishes with wavenumber as $q^2$, leading to vanishing large-scale fluctuations. 

Here, we question whether other classes of systems can generate hyperuniformity. Notably, we find that hyperuniformity can arise in a biologically realistic setting without center-of-mass conservation and without parameter fine-tuning. Inspired by a recent study of plasma cell memory, we introduce a minimal population dynamics model based on programmed cell death~\cite{bagciBistabilityApoptosisRoles2006,tysonControlCellGrowth2014,simonsTuningPlasmaCell2024}. Yet, we will show that this model captures the behavior of a broad class of systems. In this model, agents are (i) produced through a stochastic birth-type process, and are (ii) eliminated (die) when the internal state of the agent transitions away from a `viable' regime. The viable regime is (iii) stabilized when the local concentration of a shared diffusive resource exceeds a critical threshold, analogous to the role of ``survival factors'' consumed by plasma cells in the bone marrow niche. This resource, depleted by the agents, acts as the control parameter of a dynamical bifurcation that stabilizes the viable state. Together, these three ingredients define a negative feedback loop that guides the system towards a steady state where the birth rate of agents is matched by their removal (death) rate, and resource production matches its total consumption set by the population size (see Fig.~\ref{illustration}, and Fig. S1 in the Supplemental Material \cite{SeeSupplementalMaterial}). Remarkably, when the consumption rate of individuals is decreased (or the production rate is increased), the steady-state resource level tends to the critical threshold that stabilizes viability~\cite{karinEpigeneticInheritanceGene2023}. Mathematically, this model is a spatial extension of a zero-dimensional framework used recently to study biological memory in the context of the transgenerational inheritance of gene silencing~\cite{simonsTuningPlasmaCell2024}, the adaptive immune system~\cite{karinEpigeneticInheritanceGene2023}, and cell cycle control~\cite{simonsCellCycleCriticality2025}. Unlike classic models of Self-Organized Criticality \cite{bakSelforganizedCriticalityExplanation1987,watkins25YearsSelforganized2016}, the dynamics here do not exhibit large-scale avalanches. On the contrary, the system displays regular behavior over hydrodynamic length scales, allowing for systematic coarse-graining. 

Within this dynamical framework, we find that in the asymptotic limit of vanishing consumption rates, the population relaxes into a steady-state with hyperuniform spatial correlations, yet without total mass, let alone center-of-mass, conservation (as depicted in Figs.~\ref{structure}, \ref{hyperuniformityreal}, and Fig. S1 in the Supplemental Material \cite{SeeSupplementalMaterial}). Crucially, the many-body system does not undergo a phase transition. Rather, feedback drives the steady-state resource level towards its critical threshold. As this limit is approached, the resource field mediates negative-feedback interactions over a diverging range, suppressing long-wavelength density fluctuations.
 
\emph{Resource competition model} --
Formally, immobile agents (labeled $i=1,2,\dots$) are introduced stochastically in $d$-dimensional space at random positions $\mathbf{x}_i$ at a mean rate-per-unit-space volume $\lambda$. In our minimal model, agents are immobile, yet endowing them with non-zero diffusivity does not change our findings qualitatively, see Supplemental Material \cite{SeeSupplementalMaterial}. The lifetime of an individual agent $\tau_i$ is set by local environmental conditions and, in general, could be governed by a high-dimensional dynamical system. Yet, in many contexts including the one considered here, the dynamics show a critical transition between finite and diverging lifespan. In the language of dynamical systems modeling, such a transition corresponds to the elimination of a stable fixed point from some effective (possibly very high-dimensional) state space, i.e., a saddle node (SN) bifurcation. Importantly, in the vicinity of the bifurcation, the dynamics is governed by progression along a one-dimensional center manifold \cite{boxlerStochasticVersionCenter1989,hathcockReactionRatesNoisy2021},  parameterized by an effective `viability' coordinate $\nu_i\left(t\right)$, which follows the SN normal form dynamics \footnote{The Eq.~\eqref{nu} has a non-dimensional form for the viability $\nu$ and time $t$, measured with respect to a natural time and viability scale of the individual's internal dynamics. Similarly, the resource $c$  is non-dimensional by virtue of a re-scaled proportionality constant in \eqref{mu}, see Supplemental Material \cite{SeeSupplementalMaterial}.}\label{scaling} 
\begin{equation}\label{nu}
	\dot{\nu}_i=\nu_i^2 +\mu_i
\end{equation}
Here, $\mu_i=\mu\left(\mathbf{x}_i,t\right)$ serves as the bifurcation control parameter. At fixed negative $\mu<0$, the dynamics \eqref{nu} supports a stable fixed point at $\nu=-\sqrt{|\mu|}$, where the agent remains `viable'. For $\mu>0$, the fixed point is eliminated with $\nu$ monotonically increasing towards the positive half line $\nu>0$, leading to removal. The precise value of the viability at which new arrivals are initiated, set here to $\nu_0<0$, does not affect the behavior of the system (see discussion in the Supplemental Material  \cite{SeeSupplementalMaterial}). The same is true for the viability threshold for `death', set here to $\nu\geq0$. In practice, small random fluctuations, not included here, would enable noisy escape from this fixed point, leading to a finite lifetime even at $\mu<0$. While such fluctuations have a dramatic effect on the lifetime distribution \cite{simonsTuningPlasmaCell2024}, they have a sub-leading effect on density fluctuations that are at focus here, as shown in the Supplemental Material  \cite{SeeSupplementalMaterial}. Omitting them allows analytical tractability, as we discuss below. The dynamics \eqref{nu} is then coupled to a resource field $c\left(\mathbf{x},t\right)$. We denote by $c_{\text{crit}}$ the critical resource level below which the bifurcation ensues. Close to the bifurcation it is sufficient to consider a simple linear coupling \footref{scaling}\footnotemark[1]
\begin{eqnarray}\label{mu}
\mu_i\left(\mathbf{x},t\right)=c_{\text{crit}}-c\left(\mathbf{x}_i,t\right)
\end{eqnarray}
describing the elimination of the fixed point at $c<c_{\text{crit}}$. The resource is locally consumed by viable individuals, which is balanced by a fixed production rate, and is dispersed diffusively, as captured by the reaction-diffusion equation
\begin{equation}\label{y}
	\dot{c}=p\left(1- \frac{c\rho}{k}\right)+D\nabla^2c
\end{equation}  
where $p$ is the production rate, $k$ sets the ratio of production to consumption rates, and $D$ is the diffusion coefficient. The number density field of surviving agents is given by
\begin{eqnarray}\label{rho}
	\rho\left(\mathbf{x},t\right)=\sum_i\Theta\left(-\nu_i\right)\delta^d\left(\mathbf{x}-\mathbf{x}_i\right)
\end{eqnarray}
where the Heaviside theta-function $\Theta\left(\dots\right)$ selects for viable states, and the $d-$ dimensional Dirac delta-function $\delta^d\left(\dots\right)$ ensures that the consumption of resource is purely local. The coupled dynamics (\ref{nu}-\ref{rho}) describe a negative feedback, which relaxes the system towards a dynamic steady state: When resource levels are elevated above the threshold level $c>c_{\text{crit}}$, 
new agents steadily accumulate, which, according to \eqref{y}, result in increased consumption and the down-regulation of resource levels. As the system admits a unique steady state, the initial conditions affect only the transient approach to stationarity. In the asymptotic limit of vanishing individuals' consumption rate $k^{-1}\to0$, the steady state resource approaches the critical threshold $c\simeq c_{\text{crit}}$, as depicted in Fig.~\ref{illustration}. Note, however, that the strict limit of $k^{-1}=0$ is singular and does not define a regular steady state. As shown below, the particular choice of coupling to the resource (\ref{mu}-\ref{y}) also captures the generic case, where the minimal system of Eqs. (\ref{nu}-\ref{rho}) provides a universal description of critical tuning via resource competition.

\begin{figure}[]
	\begin{tabular}{ll}
	\includegraphics[width=9cm]{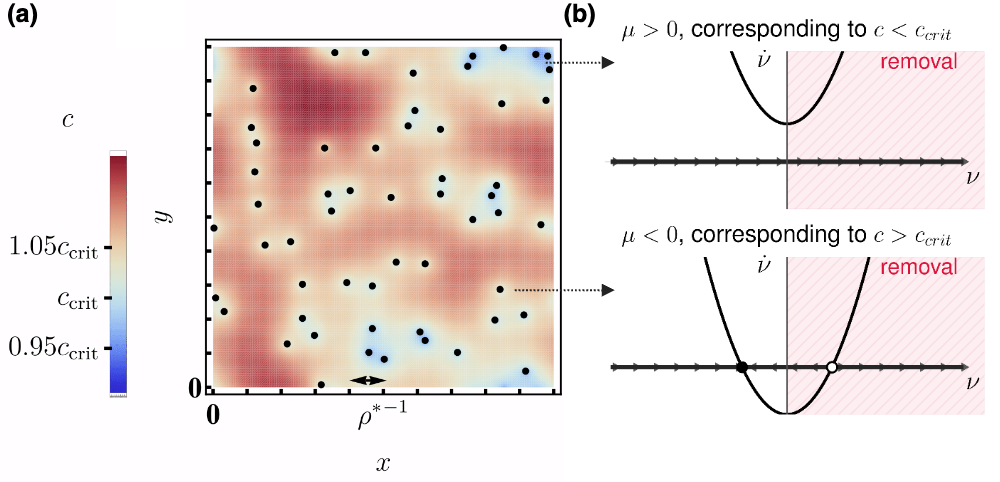}

		\end{tabular}
	\caption{(a) Snapshot of a two-dimensional realization of the population dynamics model (\ref{nu}-\ref{rho}), where particles are marked in black dots and the resource field $c$ in a color map displaying small variations around the critical value $c_{\text{crit}}=1$. (b) The instantaneous flow of the internal dynamics \eqref{nu} for two representative particles that are indicated by arrows. All parameters are set to unity except for $k=7$, corresponding to $\mu^*\simeq0.05$ \eqref{muss}.}
	\label{illustration}
\end{figure}

A corresponding `mean-field` type analysis can be found in \cite{simonsTuningPlasmaCell2024}, and is briefly recapped in the following. Remarkably, going beyond the mean-field description, we find that this dynamical steady state has hyperuniform density fluctuations as presented in Figs.~\ref{structure} and \ref{hyperuniformityreal}.

\begin{figure}[]
	\begin{tabular}{ll}	\includegraphics[width=4.1cm]{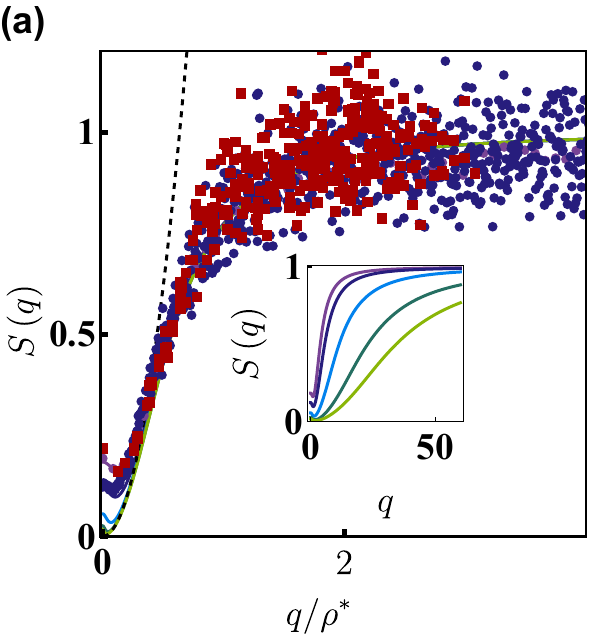}
	\includegraphics[width=4.1cm]{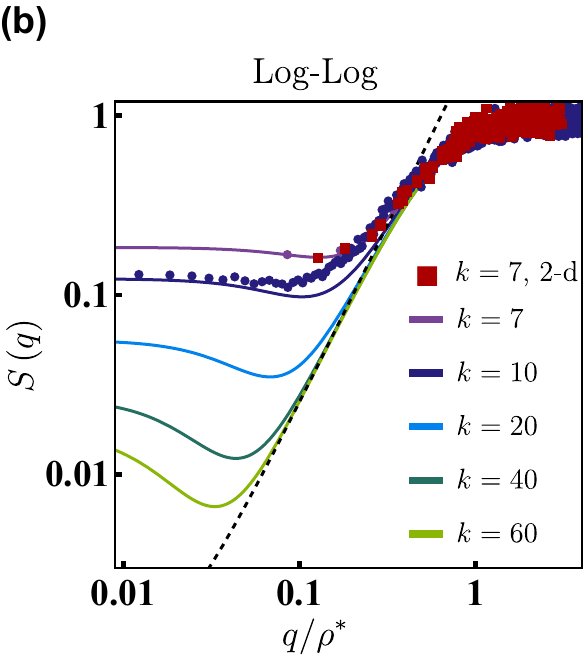}
		\end{tabular}
	\caption{Structure factor \eqref{sofq} at increasing values of $k$ approaching criticality, with all other parameters set to unity. Solid lines are the theoretical prediction from the hydrodynamic theory (\ref{yhydro}-\ref{rhohydro}) (see Supplemental Material \cite{SeeSupplementalMaterial} for their explicit expressions). Symbols denote numerical simulations of (\ref{nu}-\ref{rho}) in $1d$ (circles) and $2d$ (squares). The dashed black line is the asymptotics \eqref{smallq} at $k=60$.}
	\label{structure}
\end{figure}

\begin{figure}[]
	\begin{tabular}{ll}	\includegraphics[width=4.1cm]{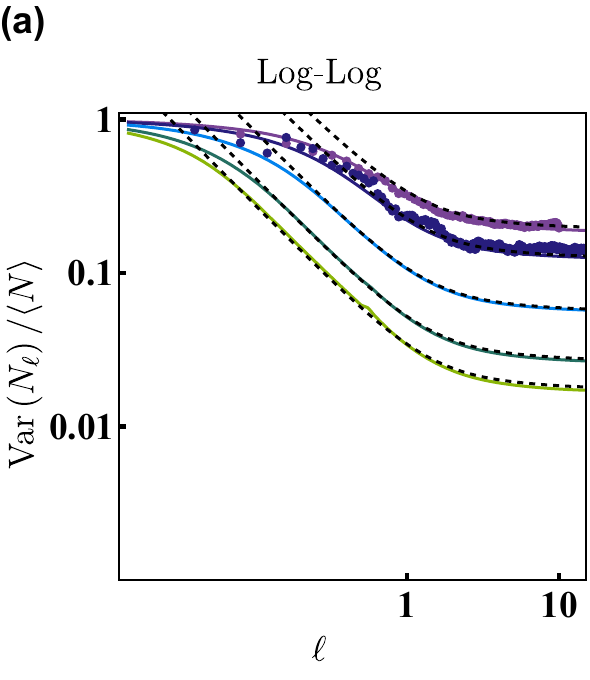}
	\includegraphics[width=4.1cm]{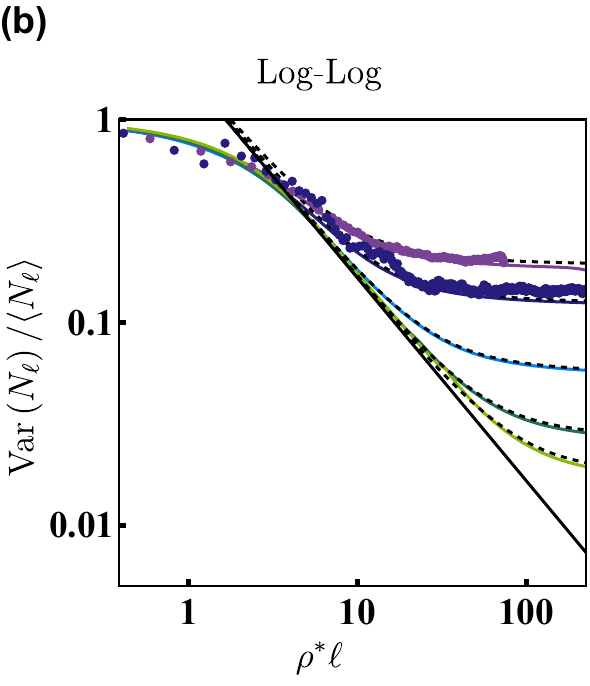}
		\end{tabular}
	\caption{Scaled number variance \eqref{hyperuniformity} color coded as in Fig. \ref{structure}, showing good agreement with the large $\ell$ asymptotics given by the first line in \eqref{largel2}, here in solid black line. Dashed black curves account for the non-perturbative corrections in the second line of \eqref{largel2}, which become dominant beyond a diverging crossover length scale $\sqrt{\ell_D\rho^*}\sim\mu^*{}^{-1/2}$.}
	\label{hyperuniformityreal}
\end{figure}

\emph{Mean-field steady-state} -- 
At the dynamic steady state, resource production matches consumption, and the agent arrival rate matches their average removal rate. At the level of mean-field, when the steady-state density and resource fields $\rho=\rho^*$ and $c=c^*$ are spatially homogeneous, these requirements are given, respectively, by 
\begin{eqnarray}\label{ss}
	 \rho^*=\frac{k}{c^*}\quad,\quad \lambda=\frac{\rho^*}{\tau\left(\mu^*\right)} 
\end{eqnarray} 
where $\lambda$ is the average arrival rate of new agents and $\tau\left(\mu^*\right)$ is the mean lifetime of agents at a fixed resource level $c^*=c_{\text{crit}}-\mu^*$. The latter can be evaluated from Eq.~\eqref{nu} as
\begin{eqnarray}\label{lifetime}
	\tau\left(\mu^*\right)=\frac{\arctan \left(|\nu_0|/\sqrt{\mu^*}\right)}{\sqrt{\mu^*}}\simeq\frac{\pi}{2\sqrt{\mu^*}}
\end{eqnarray}
with $\mu^*=c_{\text{crit}}-c^*$, and where we have expanded close to criticality at small $\mu^*$. The first equality in \eqref{ss} comes from \eqref{y}, while the second equality is simply a manifestation of Little's law in queuing theory~\cite{littleProofQueuingFormula1961}. Combining Eqs.~\eqref{ss} and \eqref{lifetime} provides the mean value of the control parameter to leading order at small $\mu^*$, with
\begin{eqnarray}\label{muss}
	\mu^*\simeq\left(\frac{\pi}{2}\frac{\lambda c_{\text{crit}}}{k}\right)^2.
\end{eqnarray} 
This expression shows that the critical resource threshold $\mu^*\to0$ is approached asymptotically  either by decreasing the average arrival rate of agents $\lambda$, or decreasing the rate of consumption $k^{-1}$. In this work, we will analyze the latter limit where the system densifies $\rho^*\sim\mu^*{}^{-1/2}$ (\ref{ss},\ref{muss}) and, as we explain below, the mean-field analysis becomes exact in any space dimension and explicit coarse-graining of the model becomes viable.

We are now in a position to evaluate the generality of our model (\ref{nu}-\ref{rho}). Little's law \eqref{ss} holds regardless of the particular choice for the dynamics of individuals \eqref{nu} or of the resource \eqref{y}; the mean density $\rho^*$ \eqref{ss} is expected to increase with decreasing particle consumption rate $k^{-1}$ for a generic resource consumption coupling. Taken together, the critical threshold $c^*\rightarrow c_{\text{crit}}$ would be approached at small $k^{-1}$ for a generic resource competition dynamics. As we show in the following, this asymptotic regime exhibits hyperuniform correlations, and these are found to be insensitive to non-linearities in the model dynamics, Eqs.~(\ref{mu}-\ref{y}). Overall, our minimal model captures generic features of self-organised hyperuniformity via resource competition.
 
\emph{Fluctuating hydrodynamics beyond mean-field} --
To analyze the hyperuniform behavior quantitatively, we turn to a coarse-grained description. This hydrodynamic limit depends crucially on the emergence of the diffusive length scale $\ell_D\equiv Dk/p\sim\mu^*{}^{-1/2}$ prescribed by the reaction-diffusion dynamics, Eq.~\eqref{y}. It represents the length scale over which the resource field $c$ varies in the presence of the $\delta$-function like consumption ``sinks'' of the agents, Eq.~\eqref{rho}. The hydrodynamic description emerges when this length scale is much larger than the average spacing between agents, $\ell_D\rho^*\gg1$, and the density field \eqref{rho} becomes smoothly varying over these scales. From Eq.~\eqref{ss}, this condition is met when $Dk^2/pc_{\text{crit}} \gg 1$, which holds in our asymptotic limit of interest, where $k$ becomes large, keeping all other parameters fixed. In this regime, nearby agents experience similar resource levels, and will have similar lifetimes, allowing for their averaging over the hydrodynamic scale. Furthermore, at coarse grained scales, the noisy arrival rate of agents becomes deterministic with weak Gaussian white-noise corrections~\cite{nisbetModellingFluctuatingPopulations2003}. These two properties allow us to write down a local and time-varying balance equation for the density of agents, akin to Little's law, that serves as a starting point for our coarse-graining procedure detailed in the Supplemental Material~\cite{SeeSupplementalMaterial}. Moreover, under weak noise, it is sufficient to consider linearized hydrodynamics for small deviations of the fields $\delta\rho=\rho\left(\mathbf{x},t\right)-\rho^*$ and $\delta c=c\left(\mathbf{x},t\right)-c^*$ around their mean-field values~\eqref{ss}. A non-trivial step in the derivation is to evaluate the agent's lifetime under time-varying resource levels. Crucially, this brings about a history dependence, which spans the agent's lifetime~\eqref{lifetime}, resulting in a system of \textit{delayed} partial differential equations
\begin{eqnarray}\label{yhydro}
	\delta\dot{ c}&=&-\frac{p}{c^*}\delta c-\frac{p}{\rho^*}\delta\rho +D\nabla^2\delta c\\\nonumber
	\delta \dot{\rho}&=&	\frac{\lambda}{\mu^*}\left[\delta c-\sqrt{\mu^*}\int_{0}^{\tau\left(\mu^*\right)}dt'\delta c\left(\mathbf{x},t-t'\right) \sin\left(2\sqrt{\mu^*}t'\right)\right]\\\label{rhohydro}
	&+&\sqrt{\lambda}\left[\xi\left(\mathbf{x},t\right)-\xi\left(\mathbf{x},t-\tau\left(\mu^*\right)\right)\right]
\end{eqnarray}   
with the unit variance Gaussian white noise $\langle\xi\left(\mathbf{x},t\right)\xi\left(\mathbf{y},t'\right)\rangle=\delta^d\left(\mathbf{x}-\mathbf{y}\right)\delta\left(t-t'\right)$.
Although the hydrodynamics (\ref{yhydro}-\ref{rhohydro}) are time non-local, they are linear, and can be integrated to yield the structure factor and number variance in terms of explicit integral expressions derived in the Supplemental Material \cite{SeeSupplementalMaterial}, showing good agreement with the results of stochastic simulation of the microscopic dynamics (Figs.~\ref{structure} and \ref{hyperuniformityreal}). Re-scaling lengths by individuals' spacing $q'=q/\rho^*$ and $\ell'=\ell\rho^*$, the structure factor approaches, non-uniformly, a limiting scaling form. 
Apart from a vanishing boundary layer around the origin, it has $\mathcal O\left(\mu^*\right)$ corrections with the small $q'$ expansion \begin{eqnarray}\label{smallq}
S\left(q\right)\simeq\frac{\pi^2}{4}\frac{D\lambda^2}{p}q'^2 +\frac{\mu^*}{c^*},
\end{eqnarray}
signaling the onset of class I hyperuniformity~\cite{torquatoHyperuniformStatesMatter2018} ($S\sim q'^2$) for any spatial dimension (see Fig.~\ref{structure}). Correspondingly, the number variance \eqref{hyperuniformity} approaches, non-uniformly, a limiting scaling form. Together with the non-perturbative large $\ell'$ corrections, it is captured by a composite matched asymptotic expansion, which in $d=1$ reads
\begin{eqnarray} \label{largel2}
&&\frac{\text{Var}\left(N_\ell\right)}{\langle N_\ell\rangle}\simeq C\sqrt{\frac{D\lambda^2}{p}}\frac{1}{\ell'}\\\nonumber
&&+\frac{\mu^*}{c^*}+\frac{2\sqrt{\mu^*}}{\pi}\frac{c^*}{p}\left[1+\frac{\sqrt{\ell_D\rho^*}}{2\ell'}\left(e^{-\frac{2\ell'}{\sqrt{\ell_D\rho^*}}}-1\right)\right]
\end{eqnarray}
with $C=0.827\dots$ provided by an explicit integral expression in the Supplemental Material  \cite{SeeSupplementalMaterial}. Equation \eqref{largel2} shows a hyperuniform scaling with the class-I  exponent $\alpha=1$ ~\cite{torquatoHyperuniformStatesMatter2018} that persists over the range $1\ll \ell'\ll \sqrt{\ell_D\rho^*}\sim\mu^*{}^{-1/2}$. The non-perturbative corrections in the second line of Eq.~\eqref{largel2} cut off the asymptotic decay past the crossover scale $\ell'\gtrsim \sqrt{\ell_D\rho^*}$, which diverges in the asymptotic limit $\mu^*\to0$  (see Fig.~\ref{hyperuniformityreal}). The same behavior holds in all space dimensions.

\emph{Analyzing hyperuniformity} -- To gain insight into the origin of hyperuniformity, we first consider the case where feedback is eliminated by holding the resource field fixed at some arbitrary profile $c^*\left(\mathbf{x}\right)<c_{\text{crit}}$ where, as shown in the Supplemental Material \cite{SeeSupplementalMaterial},  the dynamics map \textit{precisely} to a  Poisson point process, with mean density $\langle\rho\rangle=\lambda\tau\left[\mu^*\left(\mathbf{x}\right)\right]$. Such point-pattern has the non-hyperuniform  number variance $\text{Var}\left(N\right)/\langle N\rangle =1$, for all values of $\ell$. Indeed, integrating the density equation~\eqref{rhohydro} with $\delta c=0$ and vanishing initial conditions we find \footnote{We also set the noise $\xi$ to vanish in the past $t<0$, to comply with $\delta\rho(t<0)=0$}
\begin{eqnarray}\label{nofeed}
\delta\rho\left(\mathbf{x},t\right)=\sqrt{\lambda}\int_{0}^{\text{Min}\left[t,\tau\left(\mu^*\right)\right]}\xi\left(\mathbf{x},t-t'\right)dt'
\end{eqnarray}
with the uncorrelated variance $\langle\delta\rho\left(\mathbf{x},t\right)\delta\rho\left(\mathbf{y},t\right)\rangle =\langle\rho\rangle\delta^d\left(\mathbf{x}-\mathbf{y}\right)\text{Min}\left[t/\tau^*,1\right]$, coinciding with that of a Poisson process as soon as $t>\tau^*$. Importantly, close to criticality, $c^*\sim c_{\text{crit}}$, it takes a divergently long time $\tau^*\sim \mu^*{}^{-1/2}$ \eqref{lifetime} for density fluctuations to build up and saturate the Poisson statistics. This slow evolution of the density fluctuations also holds when considering the hydrodynamics (\ref{yhydro}-\ref{rhohydro}) with feedback, allowing to adiabatically eliminate the resource field \eqref{yhydro}
\begin{eqnarray}\label{css}
    \!\delta c\left(\mathbf{x}',t\right)\!\simeq\! -\frac{c^*}{\rho^*}\!\int\! \frac{d^d\mathbf {y}'}{\left(\sqrt{\ell_D\rho^*}\right)^d}G\left(\frac{|\mathbf{x}'-\mathbf{y}'|}{\sqrt{\ell_D\rho^*}}\right)\!\delta\rho\left(\mathbf{y}',t\right)
\end{eqnarray}
(see Supplemental Material \cite{SeeSupplementalMaterial}). Here we re-scaled lengths by individuals' spacing $\mathbf{x}'\equiv\mathbf{x}\rho^*$
and $G\left(\mathbf{x}\right)$ is the Green's function of the screened Poisson equation $\left(\nabla^2-1\right)G=-\delta^d\left(\mathbf{x}\right)$, which in $d=1$ takes the pure exponential form $G=e^{-|x|}/2$, and is otherwise exponentially decaying in any space dimension. The approximation \eqref{css} produces the leading term on the right hand side of the asymptotics \eqref{largel2} (see Supplemental Material \cite{SeeSupplementalMaterial} for details), and provides the mechanism for hyperuniformity: over the long individuals' lifetime, the resource has enough time to adiabatically relax over the screening length $\sqrt{\ell_D\rho^*}\sim\mu^*{}^{-1/2}$.  Inserting this approximation in the density dynamics \eqref{rhohydro}, this signifies that the resource mediates negative feedback interactions over diverging scales. The divergence of the screening length suggests an instructive analogy with the Coulomb gas, which is hyperuniform by virtue of unscreened Poisson interactions \cite{torquatoLocalDensityFluctuations2003}. However, the present dynamics remain nonconservative and time-delayed, so this analogy does not amount to an exact mapping.

A Fourier space analyses, detailed in the Supplemental Material \cite{SeeSupplementalMaterial}, provides a complementary picture and the non-perturbative corrections in the second line of Eq.~\eqref{largel2}. While, in general, density fluctuations are established over the prolonged individual's lifetime \eqref{nofeed}, large wavelength excitations $q'\ll1$ are dominated by much higher frequencies $\tau^*{}^{-1}\ll\omega$, where the time-delay hydrodynamics (\ref{yhydro}-\ref{rhohydro}) can be approximated by a time-local form. Even so, as long as the wavelengths are not too large $\mu^*{}^{1/4}\ll q'$, the relevant frequencies are not too high $\omega\ll1$, and the adiabatic approximation \eqref{css} still holds. Put together, the dynamics of wave modes $\delta\rho\left(\mathbf{q},t\right)$ with $\mu^*{}^{1/4}\ll q'\ll 1$ follows the equation
\begin{eqnarray}\label{timeloc}
   \partial_{t'}\delta\rho\simeq-\left({\frac{\pi^2}{4}\frac{D\lambda^2}{p}q'^2 +\frac{\mu^*}{c^*}}\right)^{-1}\delta\rho+\sqrt{2\rho^*}\xi\left(\mathbf{q},t'\right) 
\end{eqnarray}
where we used re-scaled time $t'=t/\tau^*$, reproducing the asymptotics \eqref{smallq} given above. This demonstrates that the mechanism for hyperuniformity here is structurally distinct from conserved-noise scenarios: the noise, being non-conservative, remains of order $\sim q^{0}$ at small $q$, while hyperuniformity is produced by a restoring kernel that diverges as 
$q^{-2}$ \eqref{timeloc}, mirroring the diverging length scales of the resource-mediated interactions \eqref{css}. At the largest wave modes $q'\ll\mu^*{}^{1/4}$, the dynamics is dominated by high temporal frequencies $\omega$, and the adiabatic approximation \eqref{css} breaks down. Accounting for these `inertial' effects in the dynamics of the resource leads to the asymptotic corrections in the second line of \eqref{largel2}, which are expected to arise generically whenever hyperuniform correlations are mediated via a diffusive resource field. 
We conclude by noting the role of the time-delay terms in \eqref{rhohydro} for hyperuniformity. These terms, which arise naturally in our model, cannot be ignored in evaluating the number variance decay \eqref{largel2}. In particular, as shown in the Supplemental Material \cite{SeeSupplementalMaterial}, they are essential to dampen small wavelength fluctuations and guarantee a bounded number variance. Nevertheless, the asymptotic hyperuniform behavior \eqref{largel2} can be reproduced starting with an alternative time-local birth-death type model that we introduce in the Supplemental Material \cite{SeeSupplementalMaterial}, where one has to impose a singular death rate.

\emph{Outlook} -- 
Hyperuniformity has been observed across diverse biological systems, from tissues \cite{kramAvianConePhotoreceptors2010,jiaoAvianPhotoreceptorPatterns2014,liuUniversalHyperuniformOrganization2024} to macroecological patterns \cite{geHiddenOrderTuring2023,huDisorderedHyperuniformitySignals2023}, yet the mechanisms driving its emergence remain unclear. We showed that hyperuniformity naturally arises in a generic class of population dynamics models, where death processes are coupled through indirect interactions between agents. Our formulation is inspired by established mechanisms of cell dynamics in tissues \cite{simonsTuningPlasmaCell2024}, yet these are general and reflect the sharp increase in agent longevity near a bifurcation. The divergence in longevity and the associated hyperuniform correlations are approached asymptotically when resource consumption rates are decreased. This differs from previous nonequilibrium settings, where hyperuniformity is reached by fine-tuning a control parameter to a critical value for a many-body phase transition ~\cite{maTheoryHyperuniformityAbsorbing2023,corteSelfOrganizedCriticalitySheared2009,hexnerNoiseDiffusionHyperuniformity2017,hexnerHyperuniformityCriticalAbsorbing2015,wieseHyperuniformityMannaModel2024}. In our model, the microscopic dynamics support a critical resource threshold where an individual's lifetime diverges, but the coupled many-body system does not exhibit a phase transition between distinct collective states. Instead, the system can only asymptotically approach this threshold, where the interaction range of the negative feedback diverges and suppresses large-scale density fluctuations. Our results, therefore, show that hyperuniformity need not be a signature of many-body criticality, and the present mechanism is complementary to, rather than belonging to, previously studied classes of nonequilibrium critical phenomena such as absorbing-state transitions. These results open a new avenue for mechanistic investigations into the origins of hyperuniformity in natural systems by identifying its underpinnings in non-conservative population dynamics.

More broadly, our dynamical framework highlights the possibility that many agents can collectively self-tune to critical states by interacting through a spatial field. Our analysis centered on the saddle-node bifurcation, a fundamental transition that captures discontinuous shifts in system dynamics. Other bifurcations have also been explored in biological contexts; for example, critical tuning to a Hopf bifurcation has been implicated in the mechanisms underlying hearing \cite{camaletAuditorySensitivityProvided2000}. Extending such frameworks to include spatial dynamics, as we have done here, may reveal new emergent phenomena with potentially important physiological consequences \cite{moraAreBiologicalSystems2011}.

\begin{acknowledgments}

\textit{Data availability}—The simulation code used to generate the numerical results in this work is publicly available \cite{wiegenfeld_resource_competition_code}.

We thank Ivan Lobaskin, Robert L. Jack, Guy Bunin, Filippo De Luca, and Xiao Ma for useful discussions. B.D.S. and T.A. are supported by the Wellcome Trust (219478/Z/19/Z) and B.D.S. by a Royal Society EP Abraham Research Professorship (RP/R/231004).
\end{acknowledgments}

\bibliographystyle{apsrev}
\bibliographystyle{iopart-num}

\bibliography{sohu}




\clearpage
\onecolumngrid

\setcounter{secnumdepth}{2}   

\setcounter{subsection}{0}
\setcounter{subsubsection}{0}
\setcounter{figure}{0}
\setcounter{equation}{0}

\renewcommand{\thesubsection}{\Alph{subsection}}
\renewcommand{\thesubsubsection}{\Alph{subsection}.\arabic{subsubsection}}

\renewcommand{\thefigure}{S\arabic{figure}}
\renewcommand{\theequation}{S\arabic{equation}}
		
		\newpage

        \graphicspath{{figures/}{SM/}}
        
		\section*{Supplemental Material to the paper ``Self-organized hyperuniformity in a minimal model of population dynamics" by T. Agranov,  N. Wiegenfeld, O. Karin and B. D. Simons}

This supplemental material serves two main purposes: First, we provide additional figures, and further details regarding the model definition and its assumptions, as well as potential extensions of the model to include the diffusion of agents and noise in the dynamics of the internal ``viability parameter'', $\nu$. These are covered in Secs.~\ref{0}, ~\ref{A}, and \ref{B}, respectively. In addition, we provide detailed derivations of some of the results of the main text.  They are covered in Secs. \ref{C}, \ref{D}, \ref{E}, \ref{F}, \ref{G}, \ref{H}, and \ref{I}. Lastly, we provide details on an alternative time-local dynamics, which emulates some of the behavior of our model in Sec. \ref{J}, and elaborate on our numerical simulations in Sec.\ref{K}.

	\subsection*{Table of contents}
	
	\begin{enumerate}[label=\Alph*.]
	{

    \item Additional figures illustrating the emergence of hyperuniformity at large $k$.
    
		\item Note on model definition: its dimensionless form, the choice of threshold value for death, and the initial condition of the state viability. 
        \item Including fluctuations in the internal viability variable $\nu$ and the addition of diffusion to the agents does not affect the main findings.
		\item Deriving the fluctuating hydrodynamics, Eq.~(11) of the main text.
        \item The feedback-less dynamics with $c\left(\mathbf{x}\right)=\text{const}$ is reduced to a Poisson point process. 
		\item Integrating the hydrodynamics to arrive at the structure factor $S$.	
		\item Arriving at a limiting scaling form of the structure factor $S$, and the small $q$ asymptotics, Eq.~(12) of the main text.

\item  Deriving the large $\ell$ asymptotics of the number variance, Eq.(13) of the main text.
\item Establishing the adiabatic limit for the resource, corresponding to the limiting scaling form of Sec.\ref{F1}.

\item Establishing the time-local approximation, which corresponds to the inner boundary expression of Sec. \ref{F2} and which leads to Eq.(16) of the main text.

\item Reproducing hyperuniform behavior within a time-local birth-death type dynamics with imposed singular death rate.

        \item  Details of numerical simulations and parameters used to produce the figures in the main text.
		}
	\end{enumerate}
		
\subsection{Additional figures illustrating the emergence of hyperuniformity in the asymptotic large-k limit}\label{0}

\begin{figure}[H]
	\begin{tabular}{ll}	
\includegraphics[width=5.4cm]{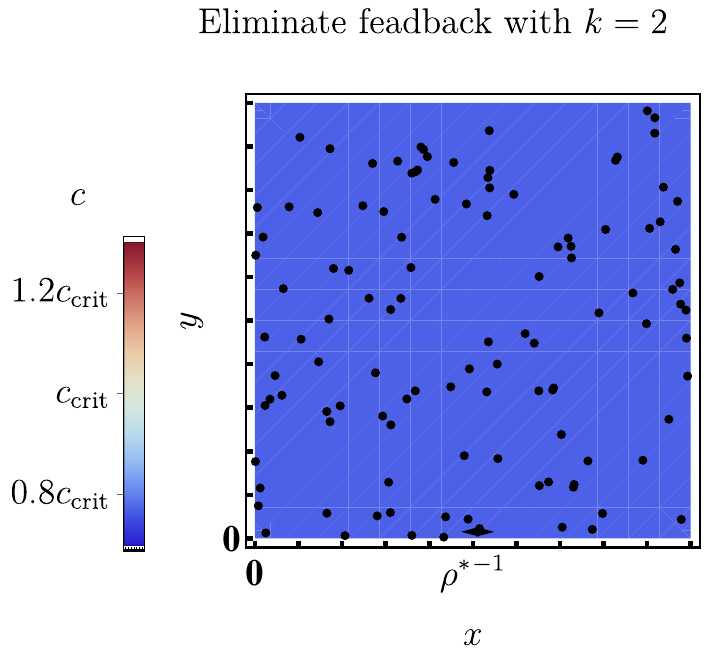}
	\includegraphics[width=4.1cm]{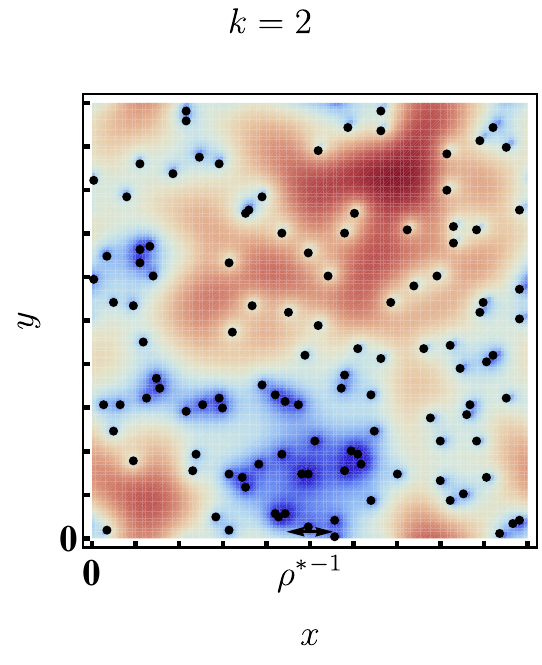}
    \includegraphics[width=4.1cm]{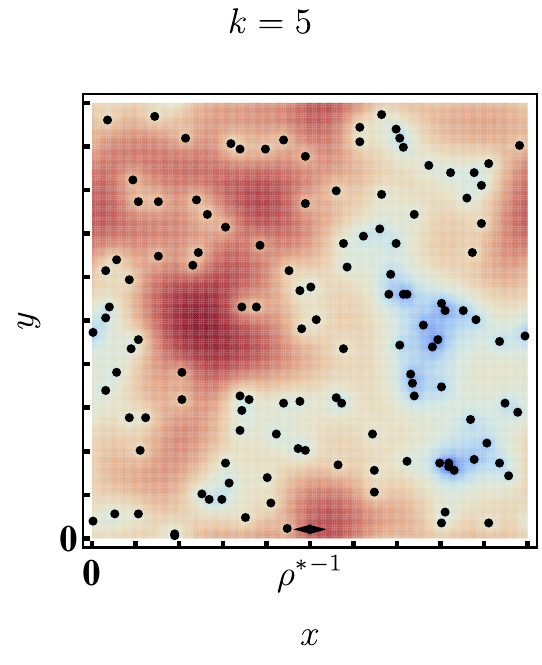}

    \includegraphics[width=4.1cm]{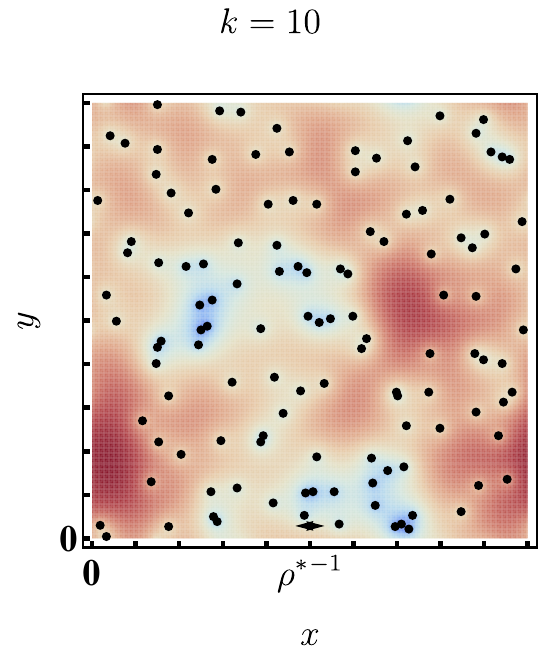}

        \\[2mm]

\hspace{0.9cm}

\includegraphics[width=4.1cm]{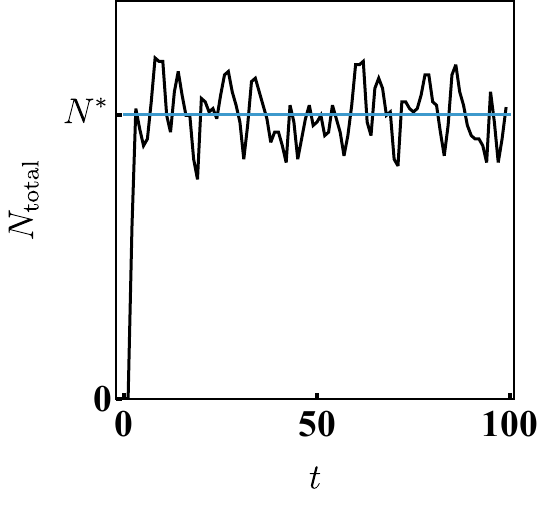}
	\includegraphics[width=4.1cm]{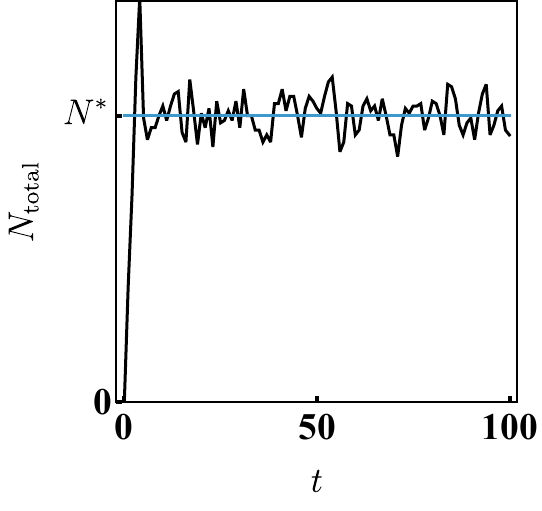}
    \includegraphics[width=4.1cm]{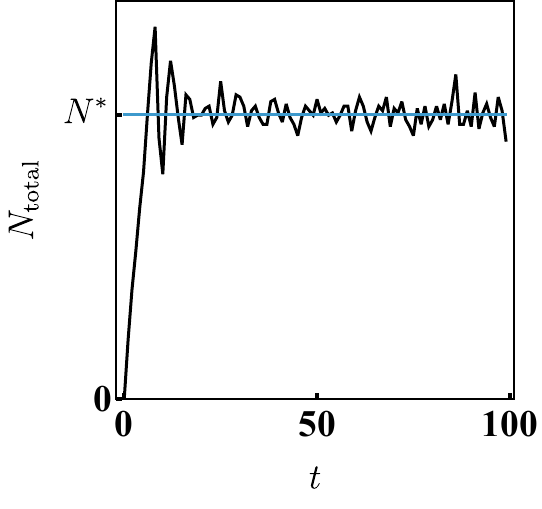}

    \includegraphics[width=4.1cm]{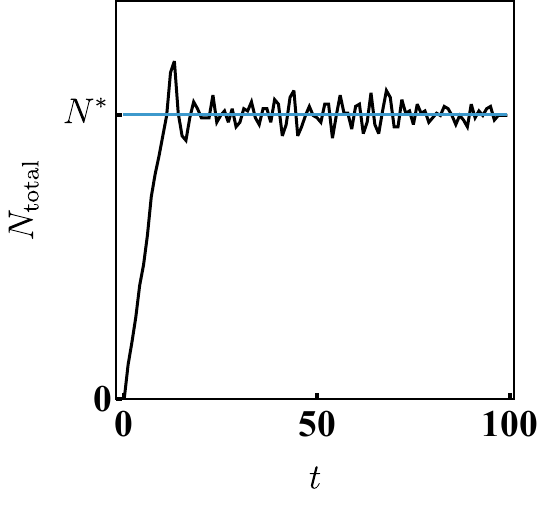}\end{tabular}
	\caption{Upper row: Steady state snapshots of the dynamics at increasing values of $k=2,5,10$. The leftmost panel corresponds to dynamics without feedback, in which the resource field is held fixed at $c^*=0.71$ (the steady-state value for $k=2$). In this case, the dynamics map exactly to a Poisson point process (see main text). The bottom row shows, for each system, the corresponding time series of total particle numbers starting from an empty occupancy. These show decreasing number fluctuations at increased values of $k$. The feedback-free case (leftmost panel) displays the largest fluctuations, consistent with Poisson statistics. In all panels, all parameters accept $k$ are set to unity. The side lengths of the $2d$ systems are set to $L=6.6, 4.8, 3.5$ for the $k=2,5,10$ systems, respectively, such that the mean total number of particles is similar $N^*\sim 120$.}
	\label{withnoise}
\end{figure}

\subsection{Note on model definition}\label{A}
The SN normal form dynamics, Eq. (3) of the main text, has a non-dimensional form.
It can always be reached, starting with dimension-full dynamics, through rescaling of coefficients. The dimension-full version reads
\begin{equation}
    \dot\nu_i = \alpha\nu_i^2 + \beta\mu
\end{equation}
with $\alpha$ and $\beta$ having dimensions of $\left[T\right]^{-1}\left[\nu\right]^{-1}$ and $\left[T\right]^{-1}\left[\nu\right]$, respectively. (Note that the bifurcation parameter $\mu$ is dimensionless.) These define natural viability and time scales $\sqrt{\beta/\alpha}$ and $1/\sqrt{\beta\alpha}$ respectively.
Rescaling the viability variable and time, $\nu \to \sqrt{\beta/\alpha}\,\nu$ and $t \to t/\sqrt{\alpha\beta}$ yields the bare form utilized in the main text. This rescaling leaves both time and the viability parameter dimensionless. Additionally, we treat the concentration field $c$ as dimensionless. This can be arranged since formally $\mu = \kappa\left(c_{crit} -c \right)$, with $\kappa$ having dimensions of inverse resource concentration, allowing the rescaling $c \to c/\kappa$.

Lastly, When introducing the model in the main text, it was claimed that the system's behavior will not depend on the initial viability $\nu_0<0$, or the viability threshold for death, which we will denote here as $\nu_{\text{death}}\geq0$. To see this, one may write the mean lifetime of agents at a fixed resource level $c^*=c_{\text{crit}}-\mu^*$ as
\begin{equation}
\tau\left(\mu^*\right) = \frac{\arctan\left({\frac{|\nu_0|}{\sqrt{\mu^*}}}\right) }{\sqrt{\mu^*}}+ \frac{\arctan\left(\frac{\nu_{\text{death}}}{\sqrt{\mu^*}}\right) }{\sqrt{\mu^*}}=\left[\frac{\pi}{2\sqrt{\mu^*}}-\frac{1}{|\nu_0|}\right]+\Theta\left(\nu_{\text{death}}\right)\left[\frac{\pi}{2\sqrt{\mu^*}}-\frac{1}{\nu_{\text{death}}}\right]+\mathcal O\left( \mu^*\right)
\end{equation}
Then, as long as $\nu_0 \sim \mathcal{O}(1)$, the $\nu_0$ dependence is sub-leading at small $\mu^*$. The dynamics have a similarly weak dependence on the cell death threshold, $\nu_{\text{death}}\geq 0$. Picking it to be positive merely results in $\tau$ doubling, which would not effect the results beyond introducing some additional constant factors.

\subsection{Including fluctuations in viability, and diffusion of the agents}\label{B}
In many biological contexts, it is natural to assume that the agents are mobile and diffusely dispersed. In addition, the dynamics of the viability parameter $\nu$ is expected to be noisy. Here, we show that both these effects, not accounted for in the main text, will not affect our main findings, namely, hyperuniformity.

\subsubsection{Accounting for diffusion in the dynamics of the agents}\label{B.1}
\begin{figure}
\includegraphics[width=0.45\linewidth]{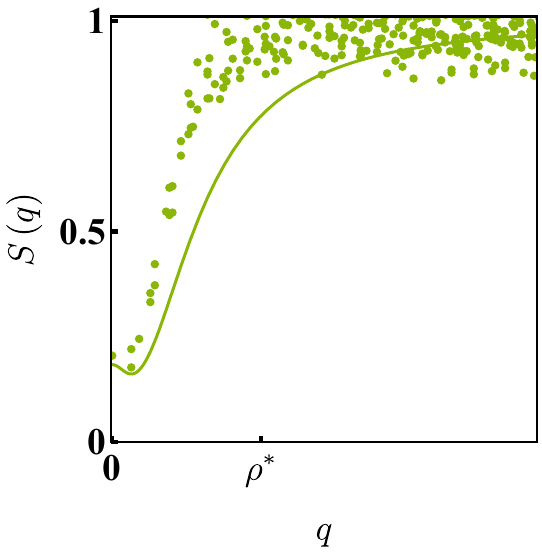}
    \caption{Structure factor $S(q)$ of the model. Data points are numerical simulations in $2d$ of the original model with the inclusion of agent diffusion with diffusion coefficient $D_{\rho}=0.05$ and with $k=7$. All other parameters are set to unity.  It displays only mild variations compared to the dynamics without diffusion (solid line). For any value of the diffusion coefficient, the hydrodynamics \eqref{yhydrodiff} predict that $S(0)$ will coincide with the value of the model without diffusion.}
    \label{diffus}
\end{figure}

Here, we show that adding diffusion to the dynamics of the agents will leave the system hyperuniform. First, we numerically simulate the microscopic dynamics of the main text, with the additional agent's diffusion, with a diffusion coefficient $D_{\rho}$. The corresponding structure factor is shown in Fig.~\eqref{diffus}. Its value at the origin $S(0)$ tends to that of the original model without diffusion,
This can be established analytically. Indeed, the diffusive displacement of agents during their lifetime can be evaluated as $\sqrt{D_{\rho}\tau^*}\sim\mu^*{}^{-1/4}\ll\ell_D$, with $\ell_D\sim\mu^*{}^{-1/2}$ the diffusive scale of the resource (see discussion in the main text). Thus, one can neglect variations in resource levels that an individual experiences due to its diffusive motion. This means that a  hydrodynamic description can be written in this case by simply supplementing the hydrodynamics of the main text with the bare diffusive terms 

\begin{eqnarray}\label{yhydrodiff}
\delta\dot{ c}&=&-\frac{p}{c^*}\delta c-\frac{p}{\rho^*}\delta\rho +D\nabla^2\delta c\\\nonumber
	\delta \dot{\rho}&=&	\frac{\lambda}{\mu^*}\left[\delta c-\sqrt{\mu^*}\int_{0}^{\tau\left(\mu^*\right)}dt'\delta c\left(\mathbf{x},t-t'\right) \sin\left(2\sqrt{\mu^*}t'\right)\right]+\sqrt{\lambda}\left[\xi\left(\mathbf{x},t\right)-\xi\left(\mathbf{x},t-\tau\left(\mu^*\right)\right)\right]\\\label{rhohydrodiff}&+&D_{\rho}\nabla ^2\delta\rho +\sqrt{2D_{\rho}}\nabla\cdot\vec{\eta}
\end{eqnarray} 
with $\vec{\eta}\left(\mathbf{x},t\right)$ a vector unit variance Gaussian white noise accounting for fluctuations in the diffusive flux. Crucially, the last two terms in \eqref{rhohydrodiff}, accounting for agent diffusion, are sub-leading $\mathcal O\left(q\right)$, compared to the non-conservative terms. Thus, they become negligible at the largest wave-modes $q\to0$.

\subsubsection{Accounting for noise in the dynamics of the viability variable $\nu$}\label{B.2}
The dynamics Eq.~(3) of the main text neglected the potential impact of fluctuations. In practice, fluctuations in the dynamics of the viability $\nu$ are expected to emerge generically from any realistic modeling of a noisy biological system. Small fluctuations can be captured by a Gaussian noise term $\xi^{\nu}_i$
\begin{equation}\label{nun}
	\dot{\nu}_i=\nu_i^2 +\mu_i+\sqrt{D_\nu}\xi^{\nu}_i\left(t\right)
\end{equation}
where the simplest case is of uncorrelated noise $\langle\xi^{\nu}_i\left(t\right)\xi^{\nu}_j\left(t'\right)\rangle=\delta_{i,j}\delta\left(t-t\right)$, and we assume small noise amplitude $D_\nu\ll1$. The effect of such fluctuations at the mean-field level has been studied in Ref.~\cite{simonsTuningPlasmaCell2024}, where they were found to have a significant impact on the lifetime distribution. Importantly, the dynamics with such fluctuations is still critically tuned, with prolonged individual lifetimes. As explained in the main text, prolonged individual lifetimes, together with the system's densification $\rho^*\sim\tau^*$, are at the basis of hyperuniformity in our model. As such, dynamical noise in \eqref{nun} is not expected to eliminate hyperuniformity. 

We could not account for the noise analytically in the spatially extended model. Our numerical simulations, presented in Fig.~\ref{withnoise}, suggest that while the form of the structure factor is manifestly deviating from the noiseless theoretical expression, it still decays at large wavelengths. We do not rule out the possibility that weak noise might terminate the structure factor decline at some small value. A more careful analysis and detailed simulations are required to determine hyperuniformity and its class here and we leave this for future investigation.

\begin{figure}[]
	\begin{tabular}{ll}	\includegraphics[width=4.1cm]{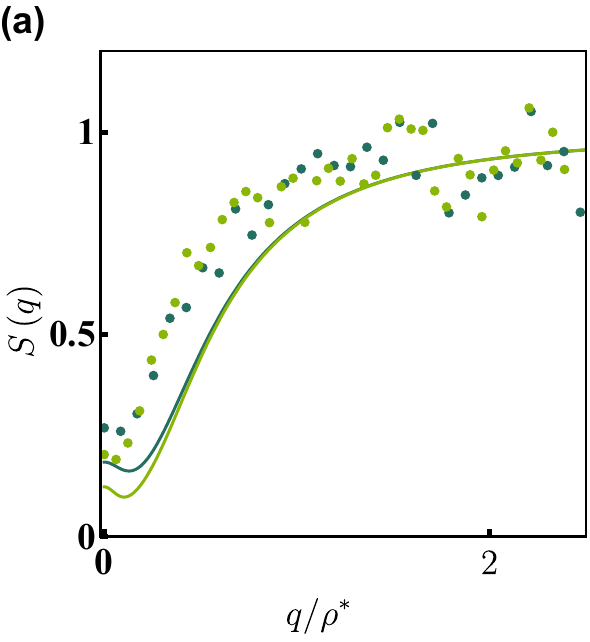}
	\includegraphics[width=4.1cm]{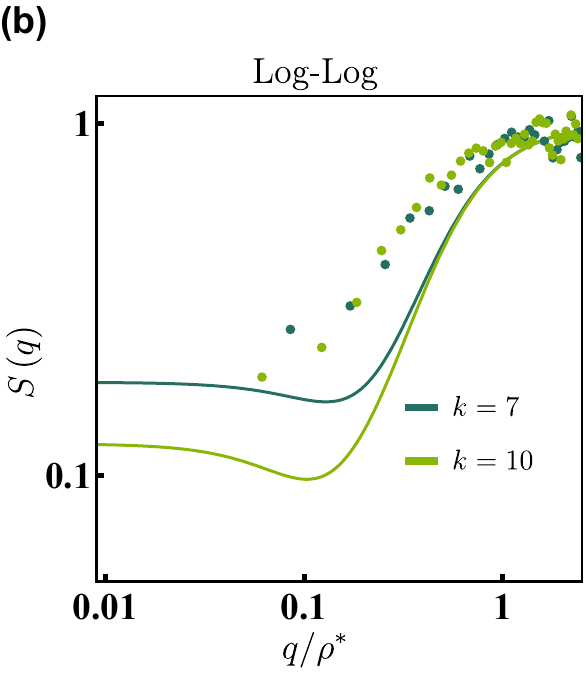}
		\end{tabular}
	\caption{Structure factor $S(q)$ of the model. Data points correspond to numerical simulations of the model with the Gaussian noise \eqref{nun} with the amplitude $D_\nu=0.01$. All other parameters, except $k=7,10$ are set to unity, as in the main text. The numerical results display deviation from the theoretical prediction for the dynamics at zero noise amplitude $D_\nu=0$, shown as solid lines. Nevertheless, the plots suggest a trend towards hyperuniformity as $k$ is increased. }
	\label{withnoise}
\end{figure}

\subsection{Derivation of the fluctuating hydrodynamics, Eq.~(11) of the main text}\label{C}
We arrived at two alternative methods to derive the fluctuating hydrodynamics. The first, which is perhaps more intuitive, is basically a balance equation for agents' arrival and removal, and can be viewed as an extension of Little's law to a time-dependent removal. A non-trivial step here is deriving the removal rate under time-varying resource levels. The second method is more formal and is based on accounting for the viability as an explicit variable of the agents' distribution $\rho=\rho\left(x,\nu\right)$, arriving at a Dean-Kawasaki type description of the dynamics. Then, integrating over the viability parameter, we arrive at the marginalized hydrodynamics Eq.~(11) of the main text. The advantages of the latter method are its versatility in analytically accounting for, e.g., noise in the dynamics of the viability. For brevity, we will only present the first method.

\subsubsection{Little's law approach}\label{C.1}
We start by writing a local version of Little's law in which we track the temporal variations in the particles' number density $\rho\left(\mathbf{x},t\right)$ in terms of local cells' loss and gain. Defining $\tau(\mathbf{x},t)$ to be the eventual lifetime of a cell initialized at location $\mathbf{x}$ at time $t$, this can be written as
\begin{equation}\label{locallittle}
    \rho(\mathbf{x},t) - \rho^* = \int_0^t\hat\lambda(\mathbf{x},t') dt' - \int_{-\infty}^tdt'\hat\lambda(\mathbf{x},t')\Theta[t'+\tau(\mathbf{x},t')]\Theta[t-t'-\tau(\mathbf{x},t')]
\end{equation}
where we have assumed that the system starts at steady-state $\rho(\mathbf{x}, 0) \equiv \rho^*$. Here $\hat\lambda(\mathbf{x},t)$ is a stochastic function modeling the full discrete, random behavior of the Poisson deposition process. It is formally given by a series of $\delta$-functions centered at the space-time locations of deposition events. The first integral is a gain term accounting for all cells deposited up to the current time. The second term integrates over all losses that occurred between the initial and current time. Notice that it has to account for the loss of cells that have been deposited at arbitrarily large times in the past $t'\in(-\infty,t)$. The $\Theta$-functions in the loss term impose the condition $0<t' +\tau(\mathbf{x},t') < t$, ensuring that it accounts for all cells that have been removed between the initial and current times. 

Over hydrodynamic scales, the noisy arrival rate $\hat\lambda(\mathbf{x},t)$ can be approximated by a fixed mean rate $\lambda$ with Gaussian white noise corrections \cite{nisbetModellingFluctuatingPopulations2003}: 
\begin{equation}\label{lambdagauss}
    \hat\lambda(\mathbf{x},t) = \lambda + \delta\lambda
\end{equation}
where $\delta\lambda(\mathbf{x},t) = \sqrt{\lambda}\xi(\mathbf{x},t)$ is the Gaussian noise
with mean $\langle\xi(\mathbf{x},t)\rangle = 0$ and variance $\langle\xi(\mathbf{x},t)\xi(\mathbf{x}',t')\rangle = \delta^d(\mathbf{x}-\mathbf{x}')\delta(t-t')$ encoding the fluctuations in the deposition rate that drive the spatiotemporal fluctuations of the system. Correspondingly, we will be focused on linearized dynamics, capturing Gaussian fluctuations around the steady state. Thus, we further define $\delta\tau(\mathbf{x},t)$ to be the variations of $\tau(\mathbf{x},t)$ around the steady-state value $\tau^* \simeq \pi/2\sqrt{\mu^*}$. Then the $\Theta$ functions inside \eqref{locallittle} may be expanded as
\begin{equation}\label{thetaexpand}  \Theta[t+\tau(\mathbf{x},t)] \simeq \Theta[t+\tau^*] +\delta\tau(\mathbf{x},t) \delta(t+\tau^*)
\end{equation} 
to first order in $\delta\tau$ since this expansion is performed under an integral sign (and similarly for the other $\Theta$ function).
Substituting the expansions \eqref{lambdagauss} and \eqref{thetaexpand} into \eqref{locallittle}, and keeping leading contributions  in small variations $\delta\lambda$ and $\delta\tau$ we obtain~\footnote{We have imposed that $\delta\tau(\mathbf{x},t <0) = 0$, and $\delta\lambda\left(t<0\right)=0$, i.e., that the system is held at steady-state until $t=0$, corresponding to the initial conditions we assumed for the density \eqref{locallittle}}:
\begin{equation}\label{locallittlelaw}
\delta\rho(\mathbf{x},t) =  \lambda\delta\tau(\mathbf{x}, t-\tau^*)+\int_{0}^{\tau^*}\delta\lambda(\mathbf{x},t-t')dt'
\end{equation}
We now turn to expressing lifetime variations $\delta\tau(\mathbf{x},t)$ in terms of the variations in the resource field $\delta c(\mathbf{x},t)$. To do so we will perturbatively solve the viability dynamics, Eq. (3) of the main text, under small, time-dependent fluctuations in the bifurcation parameter $\delta\mu(\mathbf{x},t)=-\delta c(\mathbf{x},t)$. Writing $\dot\nu_i(t) = \nu_i(t)^2 +\mu^*+\delta\mu(\mathbf{x}_i,t)$, we expand $\nu(t) = \nu^0(t) + \nu^1(t)$ with $\nu^1 \ll \nu^0$ where $\nu^0$ is the solution for $\delta\mu = 0$ and $\nu^1(t)$ is the first order correction. The lifetime variation is then expressed in terms of the viability dynamics, since for a cell deposited at time $t_i$, its viability crosses the origin at $t_i+\tau(t_i)$:

\begin{equation}
0 = \nu\Bigr|_{t=t_i+\tau(\mathbf{x}_i,t_i)} \simeq (\nu^0+\nu^1)\Bigr|_{t=t_i+\tau^*} + \dot{\nu^0}\Bigr|_{t_i+\tau^*}\delta\tau(\mathbf{x}_i,t_i) .  
\end{equation}
Thus,
\begin{equation}\label{fluclifetimeviability}
\delta\tau(\mathbf{x}_i,t_i) = - \frac{\nu^1}{\dot\nu^0}\Bigr|_{t_i+\tau^*}\end{equation}
Next, the dynamical equations for the expansion variables read
\begin{align}\label{eqnu0}
    \dot{\nu}^0 &= (\nu^0)^2 + \mu^* \\\label{eqnu1}
    \dot{\nu}^1 &= 2\nu^0(t)\nu^1(t) + \delta\mu(t)
\end{align}
with the initial conditions $\nu^0(t_i) = \nu_0$, and $\nu^1(t_i) = 0$. Then, the first equation \eqref{eqnu0} is solved by: 
\begin{equation}\label{nu0}
    \nu^0(t) = \sqrt{\mu^*}\text{tan}[\sqrt{\mu^*}(t-t_i-\tau^*)]
\end{equation}
where here, to comply with the initial condition, we use the full expression
\begin{eqnarray} \tau^*\left(\mu^*\right)=\frac{\arctan \left(|\nu_0|/\sqrt{\mu^*}\right)}{\sqrt{\mu^*}}.
\end{eqnarray}
The second equation \eqref{eqnu1} is a first-order linear ODE and can be solved with standard methods:
\begin{equation}
    \nu^1(t) = \text{exp}\left(\int^t_0 2\nu^0(t')dt'\right)\int_{t_i}^{t}dt''\text{exp}\left(-\int^{t''}_0 2\nu^0(t')dt'\right)\delta\mu(t'')
\end{equation}
Plugging in the expression for $\nu^0(t)$ \eqref{nu0} we arrive at:
\begin{equation}\label{nu1}
\nu^1(t)=\frac{\int_{t_i}^{t}dt'\text{cos}[\sqrt{\mu^*}(t'-t_i-\tau^*)]^2\delta\mu(t')}{\text{cos}[\sqrt{\mu^*}(t-t_i-\tau^*)]}\
\end{equation}
Plugging the solutions \eqref{nu0} and \eqref{nu1} into \eqref{fluclifetimeviability}, we finally arrive at the expression for the lifetime fluctuations:
\begin{equation}\label{lifetimefluc}
\delta\tau(\mathbf{x},t) = -\frac{\int_{0}^{\tau^*}dt'\cos[\sqrt{\mu^*}t']^2\delta\mu(\mathbf{x},t+\tau^*-t')}{\mu^*}    
\end{equation}
The fact that this expression is written as an integral makes it clear that the lifetime of a cell is dependent not only on $\mu$ at its time of creation, but on $\mu$ at all times throughout its life. Our linearized theory holds when $\delta\tau \ll \tau^*$. Looking at the final expression \eqref{lifetimefluc} for $\delta\tau$, it may seem at first glance that this requires $\delta\mu \ll \mu^*$. Since $\mu^*$ itself is assumed to be small, this would severely restrict the range of validity. However, notice that $\delta\mu$ enters under the integral and so its variation will tend to average out over the long integration time $\tau^*$. We therefore expect $\delta\tau \ll \tau^*$ even when $\delta\mu \sim \mu^*$. Substituting (\ref{lifetimefluc}) into (\ref{locallittlelaw}), we arrive at the desired linearized field equation:
\begin{equation}\label{rhointeq}
\delta \rho(\mathbf{x},t) = -\frac{\lambda}{\mu^*}\int_0^{\tau^*}dt'\cos(\sqrt{\mu^*}t')^2\delta\mu(\mathbf{x},t-t')+ \int_{0}^{\tau^*}\delta\lambda(\mathbf{x},t-t')dt'
\end{equation}
where the first term is a memory kernel emerging from the aforementioned dependence of cell lifetimes on the history of the field, and the second term captures fluctuations in the deposition rate. Differentiating \eqref{rhointeq} with respect to time and substituting $\delta \mu=-\delta c$ brings us to the Eq.~(11) of the main text, which is manifestly time non-local.

\subsection{The feedback-less dynamics is a Poisson process}\label{D}
Consider a general, spatially varying resource field $c^*\left(\mathbf{x}\right)<c_{\text{crit}}$. At any given time $t$, the particles found in an infinitesimal volume element $\Delta V$ around position $\mathbf{x}$ are those particles that have been deposited there within the time window $\left[t-\tau\left(\mu^*\right),t\right]$ where $\mu^*=\mu^*\left(\mathbf{x}\right)=c_{\text{crit}}-c^*\left(\mathbf{x}\right)$. As the volume element shrinks, the probability of having more than one particle in an infinitesimal volume element vanishes as $\left(\Delta V\right)^2$, and the probability $P_1$ of having one particle in this volume compartment is simply
\begin{equation}\label{p1}
    P_1=\lambda\tau\left[\mu^*\left(\mathbf{x}\right)\right]\Delta V 
\end{equation}
As particles' deposition is uncorrelated in space, so is the occupation probability of neighboring spatial compartments. Together with \eqref{p1}, this defines a spatial Poisson point process, where the probability of observing $K$ particles within any finite volume compartment $V$ follows the Poisson distribution
\begin{equation}
  P_V\left(K\right)=\frac{\bar{N}^K}{K!}e^{-\bar{N}}\quad;\quad \bar{N}\equiv\int_Vd^d\mathbf{x}\lambda\tau\left[\mu^*\left(\mathbf{x}\right)\right].
\end{equation}
This can be extended to the case of spatially and temporally varying deposition rate. 

\subsection{Deriving the structure factor, presented in Fig. 2 of the main text}\label{E}
We start with the hydrodynamic equations (10) and (11) of the main text, and perform a spatial Fourier transform. Furthermore, to facilitate comparison with mechanical oscillator systems, we  relabel the coefficients:
\begin{align}
\dot P(q,t) &= -\frac{\gamma}{M}P(q,t) - KX(q,t) \label{momentum} \\
\dot X(q,t) &= \frac{P(q,t)}{M}
- \sqrt{\mu^*} \int_0^{\tau^*} dt'\, \sin(2t'\sqrt{\mu^*}) \frac{P(q,t - t')}{M}
+ \sqrt{\lambda}\left(\xi(q, t) - \xi(q, t - \tau^*)\right)\label{position}\end{align}
where:
\[X(q,t) \equiv \delta\rho(q,t); \quad P(q,t) \equiv \delta c(q,t) ;\quad M  \equiv \frac{\mu^*}{\lambda}; \quad \gamma(q) \equiv M\left(\frac{p}{c^*}+Dq^2\right);\quad K \equiv \frac{p}{\rho^*}\]
and $\hat\xi(q,t)$ is the spatially Fourier transform of Gaussian white noise\footnote{with $\langle\hat \xi(q,t)\rangle=0$ and $\langle\hat \xi(q,t)\hat\xi(q',t')\rangle = \delta_{q, -q'}\delta(t-t')$ where the first $\delta$ is the Kronecker delta}.

Here we have connected field and density fluctuations with momentum and position, and $K, M$ and $\gamma$, provide effective spring constant, mass, and damping, respectively. This form illuminates a resemblance to a damped harmonic oscillator with the noise acting as external driving. We can then define the natural frequency $\omega_0$ and damped frequency $\omega_\gamma$ by:
\begin{equation}\label{om}
    \omega_0^2 = \frac{K}{M} \simeq \frac{2p}{\pi}\mu^*{}^{-\frac{1}{2}}; \quad \omega_\gamma = \frac{\gamma}{M} = \frac{p}{y^*}+Dq^2
\end{equation}
where we used $\rho^*=\lambda\tau^*$ with $\tau^*\simeq \pi/2\sqrt{\mu^*}$. To proceed, we Fourier transform Eqs.~(\ref{momentum}) and (\ref{position}) with respect to time, and eliminate the momentum $P$, to arrive at a closed equation for the Fourier transform of $X$.
\begin{equation}
    -i\omega \tilde X(q, \omega)+\frac{\omega_0^2\tilde X(q, \omega)}{\omega_\gamma-i\omega}= \sqrt{\lambda}\left(1-e^{i\tau^*\omega}\right)\tilde \xi(q,\omega) +\frac{2\omega_0^2\tilde X(q, \omega)}{\omega_\gamma-i\omega}\frac{e^{i\tau^*\omega}+1}{4-\frac{\omega^2}{\mu^*}}\
\end{equation}
where $\tilde\xi(q,\omega)$ is the spatially and temporally Fourier transformed Gaussian white noise \begin{eqnarray}\label{corr}
 \langle\tilde\xi(q, \omega)\tilde\xi(q', \omega')\rangle = \delta_{q, -q'} \delta(\omega+\omega')   
\end{eqnarray} and $\tilde X(\omega, q)$ is the temporal Fourier transform of $X$.
Lastly, solving for $\tilde{X}$ in terms of $\tilde{\xi}$ we arrive at the structure factor
\begin{align}\label{sq}
S\left(q\right) =  \frac{\langle\delta\rho(q,t)\delta\rho(-q, t)\rangle}{\rho^*} &= \frac{1}{\rho^*}\int 
\frac{d\omega}{2\pi} \int \frac{d\omega'}{2\pi}e^{i\left(\omega+\omega'\right)t}\langle \tilde{X}\left(q,\omega\right)\tilde{X}\left(-q,\omega'\right)\rangle \nonumber\\ &= \int _{-\infty}^{\infty} d\omega \tau^*|\zeta\left(\omega\tau^*\right)|^2 |R\left(q,\omega\right)|^2
\end{align}
with the noise and response amplitudes given, respectively, by
\begin{eqnarray}\label{zeta}
|\zeta\left(\omega\tau^*\right)|^2=\frac{1}{2\pi}\sinc^2\left(\frac{\omega\tau^*}{2}\right)
\end{eqnarray}
and 
\begin{eqnarray}\label{response}
|R|^2=\frac{\omega^2\left(\omega^2+\omega_{\gamma}^2\right)}{\left[\omega^2-\omega_0^2\frac{\sin^2\left(\frac{\omega\tau^*}{2}\right)-\frac{\omega^2\tau^*{}^2}{\pi^2}}{1-\frac{\omega^2\tau^*{}^2}{\pi^2}}\right]^2+\left[\omega\omega_{\gamma}+\frac{\omega_0^2}{2}\frac{\sin \omega\tau^*}{1-\frac{\omega^2\tau^*{}^2}{\pi^2}}\right]^2}
\end{eqnarray}
and where we used the correlator \eqref{corr}. Here the noise amplitude $|\zeta|^2$ \eqref{zeta} encodes the (Fourier transform) magnitude of the time-delayed noise $\sqrt{\lambda}\left(\xi(q, t) - \xi(q, t - \tau^*)\right)$ which play the role of the driving of the oscillator dynamics (\ref{momentum}-\ref{position}). Correspondingly, $|R|^2$ encodes the response amplitude of density fluctuations, i.e., it is coming from the homogeneous terms in the driven oscillator dynamics  (\ref{momentum}-\ref{position}).

\subsection{Deriving the limiting scaling form of the structure factor, and the small $q$ asymptotics, Eq.~(12) of the main text}\label{F}

In the following we establish small $\mu^*$ approximate expressions for the structure factor $S\left(q\right)$ \eqref{sq} in two different scaling regimes. Within the outer boundary layer $\mu^*{}^{3/8}\ll q'$, analyzed in Sec. \ref{F1}, the structure factor approaches a limiting scaling form, which corresponds to the adiabatic elimination of the resource field (see Sec.~\ref{H}). The inner boundary layer, $q'\ll1$ analyzed in Sec.\ref{F2}, corresponds to a time local approximation of the hydrodynamic equations (see Sec.~\ref{I}).

\subsubsection{Limiting scaling form at the outer boundary layer $\mu^*{}^{3/8}\ll q'$}\label{F1}
The noise amplitude~\eqref{zeta} always integrates to unity; but at large $\tau^*$, it is dominated by low frequencies $\omega\sim\tau^*{}^{-1}=2\sqrt{\mu^*}/\pi$. This mirrors the fact that close to criticality, it takes a divergently long time $\tau^*\sim \mu^*{}^{-1/2}$ for density fluctuations to build up (see discussion around Eq.(14) of the main text). Generically, this means that the leading order contribution to the integral \eqref{sq} is coming from low frequencies $\omega\sim\tau^*{}^{-1}$ where we expand the response \eqref{response}. Care is needed since the response \eqref{response} is also varying over these low frequencies. Moreover, we adopt the rescaling of space by interparticle spacing $q'=q/\rho^*$. Correspondingly, we write 
\begin{eqnarray}
|R\left(\omega,q\right)|^2=|R\left(\frac{\omega'}{\tau^*},q'\rho^*\right)|^2 
\end{eqnarray}
and expand to leading order in small $\mu^*$ to arrive at
\begin{eqnarray}\label{res}
|R|^2\simeq Q^2f\left(\omega',Q\right)\quad;\quad Q\equiv \frac{\pi^2}{4}\frac{D\lambda^2}{p}q'^2
\end{eqnarray}
with
\begin{eqnarray} f\left(\omega',Q\right)=\frac{\omega'^2}{\left[\frac{\sin^2\left(\frac{\omega'}{2}\right)-\frac{\omega'^2}{\pi^2}}{1-\frac{\omega'^2}{\pi^2}}\right]^2+\left[\omega'Q+\frac{1}{2}\frac{\sin \omega'}{1-\frac{\omega'^2}{\pi^2}}\right]^2}   
\end{eqnarray}
Put together, we arrive at a limiting scaling form for the structure factor to leading order at small $\mu^*$ (note that this expression has no $\mu^*$ dependence)
\begin{eqnarray}\label{scalelimit}
   S_{\text{outer}}\left(q\right)= Q^2\int d\omega'|\zeta\left(\omega'\right)|^2f\left(\omega',Q\right)
\end{eqnarray}
We note that this approximation corresponds to adiabatically eliminating the resource field in Eq.~(10) of the main text (see Sec.\ref{H}). In real space, this corresponds to the approximation Eq.~(15) of the main text. 

However, this approximation breaks down at too small values of $q'$ (small $Q$). To see this, we note that the characteristic scale around the origin where the integrand in \eqref{scalelimit} diverges as $\omega'\sim Q^{-1}$.  However, when arriving at \eqref{res} we neglected the term $\omega'^2/\tau^2\omega_0^2\ll1$. I.e., we assumed $\omega'\ll \mu^*{}^{-3/4}$, limiting the regime of validity of this result to be when $Q^{-1}\ll\mu^*{}^{-3/4}$, i.e., $\mu^*{}^{3/8}\ll q'$. Still, within the intermediate scaling regime $\mu^*{}^{3/8}\ll q'\ll1$ a small $q'$ expansion of \eqref{scalelimit} provides a faithful approximation of the structure factor \eqref{sq}, where for $\omega'\sim Q^{-1}\gg1$ we have
\begin{eqnarray} \label{smallq} S_{\text{outer}}\left(q'\ll1\right)\simeq Q^2\int d\omega'|\zeta\left(\omega'\right)|^2\frac{\omega'^2}{1+\omega'^2Q^2}\simeq Q=\frac{\pi^2}{4}\frac{D\lambda^2}{p}q'^2.
\end{eqnarray}

\subsubsection{Inner boundary layer covering the entire range $\ q'\ll 1$ }\label{F2}
\begin{figure}[]
	\begin{tabular}{ll}	\includegraphics[width=3.7cm]{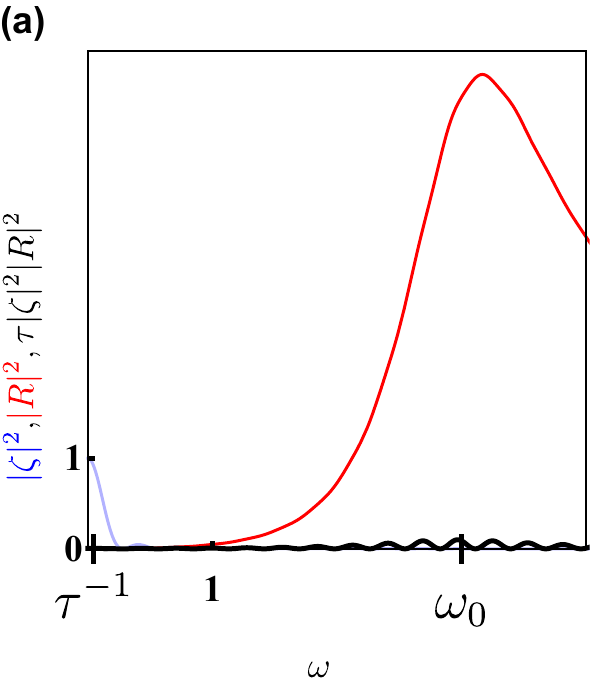}
	\includegraphics[width=3.7cm]{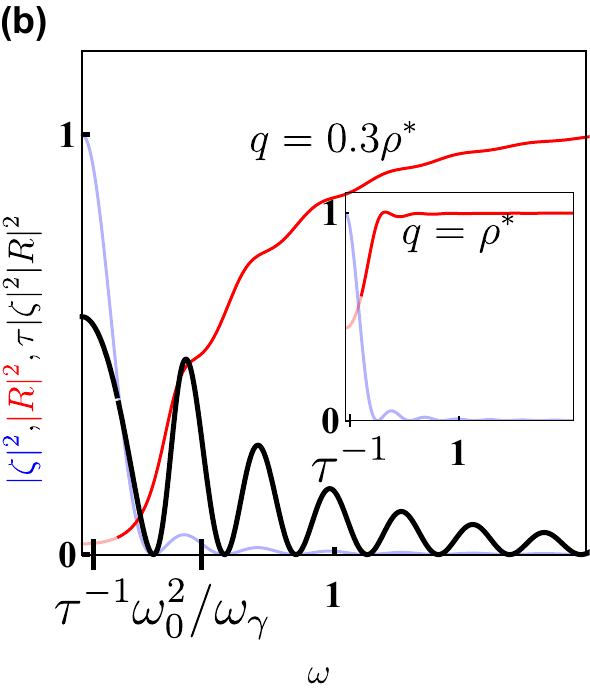}
		\end{tabular}
	\caption{The noise-response product \eqref{sq}. (a) At the longest wavelengths $q'\ll\mu^*{}^{3/8}$, the system is `under-damped', with the response peaked at the resonant frequency $\omega_0\sim\mu^*{}^{-1/4}$. (b) For long intermediate scales $\mu^*{}^{3/8}\lesssim q' \ll 1$, the noise-response product \eqref{sq} is peaked around the origin, and spans the typical frequency scale $\omega\sim\omega_0^2/\omega_\gamma\gg\tau^*{}^{-1}$. In both figures $\mu^*=0.005$.}
	\label{amp}
\end{figure}

To establish an approximation for the structure factor, we use the fact that within this entire regime, the typical frequencies which contribute to the integral \eqref{sq} are much larger than the typical noise frequency $\omega\gg\tau^*{}^{-1}$. To see this, first consider $q'\ll\mu^*{}^{3/8}$, where the system is `under-damped' $\omega_\gamma\ll\omega_0$, and the response \eqref{response} is peaked around the high frequency $\omega\sim\omega_0\sim\mu^*{}^{-1/4}$ which diverges away from the typical frequencies where the noise contributes ($\sim\tau^*{}^{-1}\sim\sqrt{\mu^*}$). This is shown in Fig.~\ref{amp}(a).
In the complementary range $\mu^*{}^{3/8}\lesssim q'\ll 1$, the system is `over-damped', and the main contribution to the integral \eqref{sq} is localized around the origin. Still, as we establish in the following, the typical frequencies that contribute to the integral scale as $\omega\sim\omega_0^2/\omega_\gamma \gg\tau^*{}^{-1}$ (see Fig.~\ref{amp}(b)). 
Thus, We have that in either scaling regimes, the relevant frequencies $\omega\gg\tau^*{}^{-1}$, and we can approximate the integrand \eqref{sq} as
\begin{eqnarray}\label{response4}
I=\tau^*|\zeta\left(\omega\tau^*\right)|^2|R|^2\simeq\frac{2}{\pi\tau^*}\frac{\omega^2+\omega_{\gamma}^2}{\left(\omega^2-\omega_0^2\right)^2+\omega^2\omega_{\gamma}^2},
\end{eqnarray}
where we replaced the rapidly oscillating $\sin^2$ in $|\zeta|^2$ by its integral over a period.
This approximation provides us with the result
\begin{eqnarray}\label{smallq2}
S_{\text{inner}}=\frac{1}{\tau\omega_\gamma}\left(1+\frac{\omega_{\gamma}^2}{\omega_0^2}\right)=\frac{2\sqrt{\mu}}{\pi}\frac{c^*}{p}\frac{1}{1+\ell_D\rho^*q'^2}+\frac{\pi^2}{4}\frac{D\lambda^2}{p}q'^2 +\frac{\mu^*}{c^*}.
\end{eqnarray}
This approximation corresponds to a time-local form for the hydrodynamic equation (11) of the main text (see Sec.\ref{I}).

To close this section, we note that the typical frequencies in the overdamped regime $\omega\sim\omega_0^2/\omega_\gamma$ can be estimated as
\begin{eqnarray*}	\omega\sim\sqrt{\frac{I\left(\omega=0\right)}{|\partial^2_{\omega}I\left(\omega=0\right)|}}=\frac{\omega_{\gamma}}{\sqrt{|2+4\left(\frac{\omega_\gamma}{\omega_0}\right)^2-2\left(\frac{\omega_\gamma}{\omega_0}\right)^4|}}
\end{eqnarray*}
 which for the over-damped regime $\omega_{\gamma}\gg\omega_0$ scale as $\omega_0^2/\omega_\gamma$.

\subsubsection{Composite matched asymptotics expansion}\label{F3}
 Notice that the regime of validity for the result \eqref{smallq2} overlaps with the regime of validity of the outer solution \eqref{scalelimit}. Correspondingly, we identify a common term in both expansions
\begin{eqnarray}
   \lim_{q'\to0} S_{\text{outer}}=\lim_{q'\to\infty} S_{\text{inner}}=\frac{\pi^2}{4}\frac{D\lambda^2}{p}q'^2
\end{eqnarray}
allowing us to write the uniform expansion, valid at small $\mu^*$ and all $q'$
\begin{eqnarray}\label{compo}
    S\left(q'\right)\simeq S_{\text{outer}}\left(q'\right)+\frac{2\sqrt{\mu}}{\pi}\frac{c^*}{p}\frac{1}{1+\ell_D\rho^*q'^2} +\frac{\mu^*}{c^*}
\end{eqnarray}
Lastly, by examining the different terms in this composite expansion, we have that within the intermediate scaling regime
\begin{eqnarray}
    \mu^*{}^{1/4}\ll q'\ll1,
\end{eqnarray}
the first two terms in the small $\mu^*$ expansion are given by
\begin{eqnarray}
    S\left(q'\right)\simeq \frac{\pi^2}{4}\frac{D\lambda^2}{p}q'^2 +\frac{\mu^*}{c^*}
\end{eqnarray}
which corresponds to Eq.~(12) of the main text. In this regime, the system both follows a time local dynamics, and the adiabatic approximation, leading to Eq. (16) of the main text (see Sec. \ref{I}).

\subsection{Deriving the large $\ell$ asymptotics of the number variance, Eq.~(13) of the main text}\label{G}
The number variance and the structure factor are related through the integral expression \cite{torquatoHyperuniformStatesMatter2018}
\begin{eqnarray}\label{var}
\frac{\text{Var}\left(N_\ell\right)}{\langle N_\ell\rangle}=\int d^d\mathbf{q} S\left(q\right)\tilde{\alpha}\left(\mathbf{q},\ell\right)
\end{eqnarray}
with $\tilde{\alpha}\left(\mathbf{q},\ell\right)$, the Fourier transform of $\alpha\left(\mathbf{r},\ell\right)$, the scaled intersection volume function of a pair of $d$-dimensional balls of radius $\ell$ separated by the radius vector $\mathbf{r}$. 
In the limit $\ell\to\infty$, where $\tilde{\alpha}$ tends to a Dirac delta function, $\delta^d\left(\mathbf{q}\right)$, we have \cite{torquatoHyperuniformStatesMatter2018}
\begin{eqnarray}\label{inf}
\lim_{\ell\to\infty}\frac{\text{Var}\left(N_\ell\right)}{\langle N_\ell\rangle}=S\left(q=0\right)=\frac{2\sqrt{\mu}}{\pi}\frac{c^*}{p}+\frac{\mu^*}{c^*}
\end{eqnarray}
where we used the boundary layer terms in \eqref{compo}. However, for not too large $1\ll\ell'\ll\sqrt{\ell_D\rho^*}$, the dominant contribution comes from the scaling form $S_{\text{outer}}$, which would give a hyperuniform decay $\ell'^{-1}$ \cite{torquatoHyperuniformStatesMatter2018}. Thus, there is an intermediate scaling regime  $\ell'\sim\sqrt{\ell_D\rho^*}$ where the $\ell'^{-1}$ decay transitions to the large $\ell'\gg\sqrt{\ell_D\rho^*}$ asymptotics \eqref{inf}, and this holds true in any space dimension.
We will derive here explicit expressions for this behavior in $1d$, although closed-form expressions are also available in higher dimensions using special functions.

Plugging \eqref{compo} in \eqref{var}, using the rescaled variables $q'=q/\rho^*,  \ell'=\ell\rho^*$ and the $1d$ expression for $\tilde{\alpha}$ \cite{torquatoHyperuniformStatesMatter2018} we have
\begin{eqnarray}\label{var2}
\frac{\text{Var}\left(N_\ell\right)}{\langle N_\ell\rangle}&=&\int_{-\infty}^{\infty} dq' \left[S_{\text{outer}}\left(q'\right)+\frac{2\sqrt{\mu}}{\pi}\frac{c^*}{p}\frac{1}{1+\ell_D\rho^*q'^2} +\frac{\mu^*}{c^*}\right]\frac{1}{\pi}\frac{\sin^2\left(q'\ell'\right)}{q'^2\ell'}\\\label{var3}
&=&\frac{\mu^*}{c^*}+\frac{2\sqrt{\mu}}{\pi}\frac{c^*}{p}\left[1+\frac{\sqrt{\ell_D\rho^*}}{2\ell'}\left(e^{-\frac{2\ell'}{\sqrt{\ell_D\rho^*}}}-1\right)\right]+\int_{-\infty}^{\infty} dq' S_{\text{outer}}\left(q'\right)\frac{1}{\pi}\frac{\sin^2\left(q'\ell'\right)}{q'^2\ell'}
\end{eqnarray}
We now aim to evaluate the contribution of the last term in \eqref{var3} coming from the scaling form $S_{\text{outer}}$. Since $S_{\text{outer}}$ is $\mathcal O\left(q'^2\right)$ \eqref{smallq}, then at large $\ell'$ it will contribute an $\ell'^{-1}$ term. Indeed, at large $\ell'$ we can safely approximate the rapidly oscillating $\sin^2$ by its integral over a period
\begin{eqnarray}
   \int_{-\infty}^{\infty} dq' S_{\text{outer}}\left(q'\right)\frac{1}{\pi}\frac{\sin^2\left(q'\ell'\right)}{q'^2\ell'}\simeq\frac{1}{\ell'}\times\frac{1}{2\pi}\int_{-\infty}^{\infty} dq' \frac{S_{\text{outer}}\left(q'\right)}{q'^2} 
\end{eqnarray}
Plugging the expression \eqref{scalelimit}, we have
\begin{eqnarray}
    \frac{1}{2\pi}\int_{-\infty}^{\infty} dq' \frac{S_{\text{outer}}\left(q'\right)}{q'^2} =\frac{1}{\pi^2}\int_{-\infty}^{\infty} d\omega'\sin^2\left(\frac{\omega'}{2}\right)\int_{-\infty}^{\infty}dq'\frac{Q^2\left(q'\right)/q'^2}{a_1^2\left(\omega'\right)+\left[\omega'Q\left(q'\right)+a_2\left(\omega'\right)\right]^2}
\end{eqnarray}
with
\begin{eqnarray}    a_1=\frac{\sin^2\left(\frac{\omega'}{2}\right)-\frac{\omega'^2}{\pi^2}}{1-\frac{\omega'^2}{\pi^2}}\quad;\quad a_2=\frac{1}{2}\frac{\sin \omega'}{1-\frac{\omega'^2}{\pi^2}}\quad;\quad Q=\frac{\pi^2}{4}\frac{D\lambda^2}{p}q'^2
\end{eqnarray}
Then, rescaling 
\begin{eqnarray}
    q''=\sqrt{Q\left(q'\right)}
\end{eqnarray}
we arrive at (omitting primes)
\begin{eqnarray}
    \frac{1}{2\pi \ell'}\int_{-\infty}^{\infty} dq' \frac{S_{\text{outer}}\left(q'\right)}{q'^2} &=&\frac{1}{2\pi \ell'}\sqrt{\frac{\lambda^2D}{p}}\int_{-\infty}^{\infty}d\omega'\sin^2\left(\frac{\omega'}{2}\right)\int_{-\infty}^{\infty}dq\frac{q^2}{a_1^2\left(\frac{\omega'}{2}\right)+\left[\omega'q^2+a_2\left(\omega'\right)\right]^2}\\\label{complex}
    &=&\frac{1}{2\pi\ell'}\sqrt{\frac{\lambda^2D}{p}}\int_{-\infty}^{\infty} d\omega'\frac{\sin^2\left(\frac{\omega'}{2}\right)}{\omega'{}^2}\int_{-\infty}^{\infty}dq\frac{q^2}{\left[q-\sqrt{x_1}\right]\left[q+\sqrt{x_1}\right]\left[q-\sqrt{x_2}\right]\left[q+\sqrt{x_2}\right]}
\end{eqnarray}
with
\begin{eqnarray}
    x_{1,2}=-\frac{a_2}{\omega}\pm i\frac{|a_1|}{|\omega|}.
\end{eqnarray}
The $q$ integration in \eqref{complex} can be performed by completing the integration path into a closed contour in the complex plane, e.g, via the arc $|q|\to\infty$ in the upper half complex plane, where we finally arrive at

\begin{eqnarray}\label{largel0}
 \frac{1}{2\pi \ell'}\int_{-\infty}^{\infty} dq' \frac{S_{\text{outer}}\left(q'\right)}{q'^2} =C\sqrt{\frac{\lambda^2D}{p}}\frac{1}{\ell'}\quad;\quad C=\frac{1}{2\sqrt{2}}\int d\omega'\frac{\sin^2\left(\frac{\omega'}{2}\right)}{\omega'{}^2}\frac{1}{\sqrt{\sqrt{\frac{a_1^2\left(\omega'\right)}{\omega^2}+\frac{a_2^2\left(\omega'\right)}{\omega^2}}+\frac{a_2\left(\omega'\right)}{\omega}}}\simeq 0.8276
\end{eqnarray}
which together with \eqref{var3} provides Eq.~(13) of the main text.

\subsection{Establishing the adiabatic limit for the resource, corresponding to the limiting scaling form \eqref{scalelimit}  of Sec.\ref{F1}}\label{H}
The approximation \eqref{res}, which provides the limiting scaling form \eqref{scalelimit}, corresponds to the adiabatic elimination of the resource. We now see how this approximation emerges directly from the hydrodynamics, Eqs. (10-11) of the main text. To observe this, we rescale time by the particle's lifetime $t'\equiv t/\tau$, and the resource by the prefactor $V=\frac{\rho^*}{\mu^*}\delta c$ to arrive at
\begin{eqnarray}\label{yd}
	\partial_{t'} V&=&-\tau \left(\frac{p}{c^*}V-D\nabla^2V+\frac{p}{\mu^*}\delta\rho  \right)\\\label{rhod}
	\partial_{t'}\delta\rho&=&	V-\frac{\pi}{2}\int_{0}^{1}d\tau'V\left(t'-\tau'\right) \sin\left(\pi\tau'\right)+\sqrt{\rho^*}\left[\xi\left(x,t'\right)-\xi\left(x,t'-1\right)\right]
\end{eqnarray}
Within these variables, the equation for density variations $\delta\rho$ has $\mathcal O\left(1\right)$ terms, with a driving noise whose variance equals the mean density. In contrast, the $V$ equation \eqref{yd} has the large $\tau$ prefactor, where density variations $\delta\rho$ act as a driving term. Thus, unless the driving term $\delta\rho$ is rapidly oscillating in time, this allows for the adiabatic elimination of $V$
\begin{equation}\label{ad}
    \tau\left(\frac{p}{c^*}V-D\nabla^2V+\frac{p}{\mu^*}\delta\rho\right)\simeq 0,
\end{equation}
whose solution in terms of the resource is given by Eq.~(15) of the main text.

As we saw in Sec.~\ref {F2}, this approximation breaks down when examining large wavelength modes, where typical excitations of the density are characterized by fast temporal frequencies. Within the time rescaling used above and in Sec.\ref{F2}, the approximation \eqref{ad} breaks once $\omega'\sim Q^{-1}\gg\tau$, and the time derivative in \eqref{yd} overwhelms the $\tau$ prefactor on the right-hand side. Within space Fourier variables we must have $\mu^*{}^{1/4}\ll q'$ for the approximation to hold. Notice, however, that this condition is more stringent then the one of Sec.\ref{F1}. The reason is that it also accounts for the sub-leading term $p/c^*V$. 
Indeed, notice that within scaled space variables $\mathbf{x}'=\rho^*\mathbf{x}$
\begin{eqnarray}
 D\nabla^2V =\frac{\pi^2}{4}\frac{D\lambda^2}{\mu^*}\nabla_{x'}^2V.
\end{eqnarray}
Correspondingly, to leading order, the adiabatic approximation continues to hold at even smaller values of $q'$ where it is now reduced to the Poisson equation
\begin{eqnarray}
        \tau\left(-\frac{\pi^2}{4}\frac{D\lambda^2}{\mu^*}\nabla_{x'}^2V+\frac{p}{\mu^*}\delta\rho\right)\simeq 0
\end{eqnarray}
and this holds for
$\omega'\sim Q^{-1}\ll\left(\tau/\mu^*\right)q'^2$, i.e., $\mu^*{}^{3/8}\ll q'$, in agreement with the analyses of Sec.\ref{F1}.

\subsection{Establishing the time-local approximation, which corresponds to the inner boundary expression of Sec. \ref{F2} and which leads to Eq.~(16) of the main text}\label{I}
The approximation \eqref{response4}, which leads to the inner boundary expression \eqref{smallq2}, corresponds to a time local approximation of the hydrodynamics, which we now turn to derive.
As shown in Sec.\ref{F3}, long wavelength excitations $q'\ll1$ are characterized by fast frequencies $\omega\gg\tau^*{}^{-1}\sim\sqrt{\mu^*}$. For such fast temporal oscillations, the time integral in Eq.~(11)
\begin{eqnarray}
\sqrt{\mu^*}\int_{0}^{\tau\left(\mu^*\right)}dt'\delta c\left(\mathbf{x},t-t'\right) \sin\left(2\sqrt{\mu^*}t'\right)\sim \frac{\mu^*}{\omega^2}
\end{eqnarray}
is negligible. Next, we denote the Gaussian noise term in Eq. (11) of the main text as $\Xi\left(\mathbf{x},t\right)\equiv \xi\left(\mathbf{x},t\right)-\xi\left(\mathbf{x},t-\tau^*\right)$ having the two-point function
\begin{eqnarray} \label{bxi}  \langle\Xi\left(\mathbf{x},t\right)\Xi\left(\mathbf{y},t'\right)\rangle=2\delta^d\left(\mathbf{x}-\mathbf{y}\right)\delta\left(t-t'\right)-\delta^d\left(\mathbf{x}-\mathbf{y}\right)\left[\delta\left(t-t'-\tau^*\right)+\delta\left(t-t'+\tau^*\right)\right]
\end{eqnarray}
The last two cross-terms in \eqref{bxi} capture correlations between distant time points separated by $\tau^*$. These are negligible for high frequency modes $\omega\gg\tau^*{}^{-1}$. Indeed, in the time Fourier domain, they contribute the rapidly oscillating $\sim\cos\left(\omega\tau^*\right)$ term of the noise amplitude, which integrates to zero.  
Taken together, we find that long wavelength excitations $\delta c\left(q',t\right)$ and $\delta \rho\left(q',t\right)$ with $q'\ll1$ are described by the time local dynamics
\begin{eqnarray}\label{yhydro2}
	\delta\dot{ c}&=&-\frac{p}{c^*}\delta c-\frac{p}{\rho^*}\delta\rho -D\rho^*{}^2q'^2\delta c\\\label{rhohydro2}
	\delta \dot{\rho}&=&	\frac{\lambda}{\mu^*}\delta c+\sqrt{2\lambda}\xi\left(\mathbf{q}',t\right)
\end{eqnarray}  
These coupled equations provide the structure factor \eqref{smallq2}. Note that this structure factor has unbounded small wavelength fluctuations  $S\sim q'^2$ as $q'\to\infty$, indicating that the time delay terms are crucial to bound these small-scale fluctuations, and arrive at a bounded number variance.

Lastly, as explained in Secs.\ref{F1}, for not too large wavelengths $\mu^*{}^{3/8}\ll q'$, the resource can be adiabatically eliminated. If, moreover, $\mu^*{}^{1/4}\ll q'$, we can include sub-leading corrections (see Sec.\ref{F3} and \ref{H}) where it is approximated by
\begin{eqnarray}
    \delta c\simeq -\frac{c^*}{\rho^*}\frac{\delta\rho}{1+\ell_D\rho^*q'^2}
\end{eqnarray}
which, when inserted into \eqref{rhohydro2}, provides Eq.~(16) of the main text.

\subsection{Time-local birth-death dynamics with imposed singular death rate}\label{J}
The hyperuniform behavior can be reproduced within a time-local dynamics of a birth-death type. I.e., we propose a simplified model where randomly deposited individuals undergo death (and elimination from the system) at a resource-dependent Poisson rate $\mathcal D$. To emulate the critical behavior of the saddle noise dynamics, one has to impose a singular death rate, which vanishes at a critical value of the resource
\begin{eqnarray}
    \mathcal D\left(\mu\right)=\frac{1}{\tau\left(\mu\right)}\Theta\left(\mu\right)=\frac{2\sqrt{\mu}}{\pi}\Theta\left(\mu\right)\quad;\quad \mu=c_{\text{crit}}-c\left(\mathbf{x}\right)
\end{eqnarray}
where we used the expression for $\tau$ given by Eq. (8) of the main text. Adopting the same hydrodynamic limit as in the main text, the fluctuating hydrodynamic description of this model reads
\begin{eqnarray}\label{cloc}
	\dot{c}&=&p\left(1-\frac{c\rho}{k}\right)+D\nabla^2c\\\label{rholoc}
    \dot{\rho}&=&\lambda-\frac{\rho}{\tau\left(\mu\right)}+\sqrt{\lambda}\xi+\sqrt{\frac{\rho}{\tau\left(\mu\right)}}\eta
\end{eqnarray}
with $\eta$ and $\xi$ uncorrelated $\langle \eta \xi\rangle=0$ unit variance Gaussian white noise terms. The additional $\eta$ noise term is coming from the stochasticity in the Poisson death events \cite{nisbetModellingFluctuatingPopulations2003}. 

The steady state, given by stationarity of (\ref{cloc}) and (\ref{rholoc}), takes the same form as Little's law, Eq. (7) of the main text
\begin{equation}
\rho^*=\frac{k}{y^*}\quad;\quad
\lambda=\frac{\rho^*}{\tau\left(\mu^*\right)}
\end{equation}
Indeed, the mean particle's lifetime for this Poisson death process is given by the inverse death rate.

In the hydrodynamic limit, the structure factor is found within linearized hydrodynamics, which takes a very similar form to (\ref{yhydro2}) and (\ref{rhohydro2}),
\begin{eqnarray}\label{ylin}
	\delta\dot{c}&=&-\frac{p}{c^*}\delta c-\frac{p}{\rho^*}\delta\rho +D\nabla^2\delta c\\\label{rholin}
        \delta\dot{\rho}&=&-\lambda\frac{\delta\rho}{\rho^*}+\frac{1}{2}\frac{\lambda}{\mu^*}\delta c+\sqrt{2\lambda}\xi
\end{eqnarray}
with the crucial difference being the additional decay term $-\lambda \delta\rho/\rho^*$ in \eqref{rholin}. This term is essential to bound small wavelength fluctuations and arrive at a finite structure factor. In the original model, this is achieved via the time delay terms in Eq. (11) of the main text, establishing their crucial role for hyperuniformity in this model. 

For the coupled system, Eqs.~(\ref{ylin}) and (\ref{rholin}), one can derive the structure factor explicitly as
\begin{eqnarray}
    S\left(q\right)=\frac{1}{\omega_\gamma\tau+1}+\frac{\omega_\gamma^2\tau^2}{\left(\frac{1}{2}\omega_0^2\tau^2+\omega_\gamma\tau\right)\left(\omega_{\gamma}\tau+1\right)}
\end{eqnarray}
with $\omega_0$ and $\omega_\gamma$ defined in \eqref{om}. As for the original model, this structure factor approaches non-uniformly a limiting hyperuniform scaling form. Apart from a vanishing boundary layer near the origin, it has the small $q'$ expansion
\begin{eqnarray}\label{smallq}
S\left(q\right)\simeq \frac{\pi^2}{2}\frac{D\lambda^2}{p}q'^2 +2\frac{\mu^*}{c^*},
\end{eqnarray}
which, up to a factor of $2$, coincides with the original model, Eq. (12) of the main text.
\subsection{Details of numerical simulations and parameters used to produce the figures in the main text}\label{K}
In order to perform simulations of the model, we employed a simple finite element spatial discretization. In doing so, space is replaced by a lattice with spacing $dx$, and the system is incrementally updated in time steps of $dt$. In order to maintain numerical stability of the diffusive evolution of the resource field on such a discrete lattice, $dt$ must scale as $\frac{Cdx^2}{D}$ where $D$ is diffusion and $C$ is some constant. In our numerical simulations, we chose to approach criticality by increasing $k \to  \infty$ and holding $p=c_{crit} = \lambda=D=\mathcal{O}(1)$. Since $\rho^* \approx \frac{k}{c_{crit}}$, this increases the number of cells in the system and thus the required spatial resolution of the simulations. Schematically, given $dx = \mathcal{O}(k^{-1})$, this means that simulation runtime scales as $T \sim (dx^ddt)^{-1} \sim\mathcal{O}(k^{d+2})$ where $d$ is the spatial dimension.
We found that as we reduced $dx$, the simulated and theoretical values of $\mu^*$ converged, and that taking $dx$ to be 10-15 times smaller than $\frac{1}{\rho^*}$ was sufficient to yield good agreement in $\mu^*$. The simulation code can be found at \url{https://github.com/natan-wiegenfeld/Resource-Competition-Hyperuniformity-Simulations} \cite{wiegenfeld_resource_competition_code}.

\end{document}